\documentclass[artical,3p,times]{elsarticle}
\usepackage{lineno,hyperref,graphicx}
\usepackage{multirow}
\usepackage[table,xcdraw]{xcolor}
\usepackage{caption}
\usepackage{booktabs}
\usepackage{floatrow}
\usepackage{amsmath}
\usepackage{color}
\journal{Journal of \LaTeX\ Templates}
\hypersetup{colorlinks,% Reference color setup	
	linkcolor=Cerulean,%
	citecolor=Cerulean}

\begin{document}

\begin{frontmatter}

\title{Maximizing spreading in complex networks with risk in node activation}
\tnotetext[mytitlenote]{Fully documented templates are available in the elsarticle package on \href{http://www.ctan.org/tex-archive/macros/latex/contrib/elsarticle}{CTAN}.}

\author[mymainaddress,mymainaddress2]{Leyang Xue}

\author[mymainaddress]{Peng Zhang}

\author[mysecondaryaddress]{An Zeng\corref{mycorrespondingauthor}}
\cortext[mycorrespondingauthor]{Correspondence and requests for materials should be addressed to A.Z}
\ead{anzeng@bnu.edu.cn}

\address[mymainaddress]{School of Science, Beijing University of Posts and Telecommunications, Beijing 100876, China.}
\address[mymainaddress2]{International Academic Center of Complex Systems, Beijing Normal University, Zhuhai 519087, China.}
\address[mysecondaryaddress]{School of Systems Science, Beijing Normal University, Beijing 100875, China.}

\begin{abstract}

It is widely acknowledged that the initial spreaders play an important role for the wide spreading of information in complex networks. Thus, a variety of centrality-based methods have been proposed to identify the most influential spreaders. However, most of the existing studies have overlooked the fact that in real social networks it is more costly and difficult to convince influential individuals to act as initial spreaders, resulting in a high risk in maximizing the spreading. In this paper, we address this problem on the basis of the assumption that large-degree nodes are activated with a higher risk than small-degree nodes. We aim to identify the effective initial spreaders to maximize spreading when considering both the activation risk and the outbreak size of initial spreaders. On random networks, the analytical analysis reveals that the degree of optimal initial spreaders does not correspond to the largest degree of nodes in the network but rather be determined by infection probability and difference of activation risk among nodes with different degree. Here, we propose a risk-aware metric to identify the effective spreaders on real networks. The numerical simulation shows that the risk-aware metric outperforms the existing benchmark centralities in maximizing the effective spreading.
\end{abstract}


\end{frontmatter}

%\linenumbers

\section{Introduction}

Social networks as a media play an irreplaceable role in the spreading of information, opinion, ideas, innovation, and rumors\cite{centola2010spread, montanari2010spread,yang2020containment}. In social networks, identifying influential spreaders could help us efficiently control the outbreak of epidemics\cite{pastor2002immunization}, successfully advertise a new product\cite{opuszko2013effects}, and largely facilitate information dissemination\cite{kimura2007extracting}. A recent example is that super spreading events (SSE) are associated with explosive growth early in an outbreak of COVID-19\cite{frieden2020identifying}, identifying the high-risk setting of SSE and timely implementation of interventions can help prevent and control future infections disease outbreaks. Hence, the problem of influence maximization has received extensive attention from multiple disciplines, such as mathematics, physics, computer science, sociology\cite{banerjee2018survey,li2018influence,morone2015influence,lu2016vital,aral2018social}.  

%The problem of selecting the set of initial nodes as source spreaders to achieve maximum scale of spreading is defined influence maximum problem (IMP). The research on IMP initially originated from the virtual market that is a business strategy by considering existing social network to promote a product \cite{richardson2002mining}. Lately, the IMP was confirmed as a NP-hard problem and is exactly solvable for very small network by greedy optimization algorithms \cite{kempe2003maximizing}. But for a large-scale social network, choosing the top-ranking influential nodes as source spreaders is regarded as a classical and feasible strategy, which is called centrality-based heuristics algorithms.

Thus far, large effort has been devoted to identifying influential nodes\cite{lu2011leaders,chen2012identifying, chen2013identifying,de2014role, zhang2016identifying}. Originally, some well-known centralities are used to identify the influential nodes in complex networks, such as degree, closeness\cite{sabidussi1966centrality}, betweenness\cite{freeman1978centrality}, eigenvector\cite{bonacich2007some}, Katz\cite{zhan2017identification} and subgraph\cite{estrada2005subgraph}. Then, Kitsak et al.\cite{kitsak2010identification} argue that the location of nodes in the network is more important than its immediate neighbors in evaluating the spreading influence of nodes, and use the k-shell decomposition to measure the location of nodes in the network. If the node is located in the core of the network, it will have a higher influence to disseminate the information than the one located in the periphery. Although some nodes with the same coreness are indistinguishable in some cases (e.g. Barabasi-Albert network and tree-like network) by k-shell because it is highly coarse-grained, their findings have been widely disseminated and draw attention to the problem. Subsequently, many new methods have been devised to identify the influential spreaders\cite{erkol2019systematic}. For example, Zeng et al. present a mixed degree decomposition (MDD) procedure to rank the spreader by considering the residual degree and exhausted degree (In the $k_s$ layer decomposition, nodes with degree smaller than the $k_s$ value of the current layer are successively pruned until no more such nodes remain. In this case, for each remaining node, residual degree refers to the number of links connecting to other remaining nodes, and exhausted degree denotes the number of links connecting to removed nodes), and show that the MDD outperforms the k-shell\cite{zeng2013ranking}; L\"u et al. find that H-index could better quantify the node influence than the degree and coreness (i.e. k-shell)\cite{lu2016h}; Zareie et al. propose an improved cluster rank approach to identify the influential nodes by taking into account the common hierarchy of nodes and their neighborhood set\cite{zareie2020finding}.

The main implicit assumption in most relevant studies is that the probability of influential individuals to act as initial spreaders is independent of personal influence (i.e. status level in the social network, referred to the rank, deference, or popularity). But when dealing with a real application of influence maximization, we recognize that some realistic factors (e.g. the cost, accessibility of influential individuals) still need to be considered into the problem. In related work, many researchers propose a variety of methods to identify individuals with high influence and suggest that marketing managers should try to seed them to promote products on online social networks. The high-influence seeding targets indeed have higher returns but associated with high risks due to the following reason. $(1)$ There is a constrain in the budget of promotional activities. In marketing,  hiring higher influence individuals to promote the products need pay higher cost than average people. $(2)$ Even though marketing managers could pay for the high cost, celebrities are not willing to promote products because of duty or time constrain\cite{bakshy2011everyone}, resulting in a high risk in maximizing the spreading. The related empirical study also confirms that the probability of responding to an endorsement request is dependent on status, and it sharply declines with the status difference in a user-generated network\cite{lanz2019climb}. To make the problem closer to the real situation, we relax the assumption and capture reality factors as the activation risk of initial spreaders. Here, we assume that individuals with higher influence tend to be activated (i.e. be accepted to act as initial spreaders) at a higher risk than ordinary. When one takes into account both the outbreak size and the activation risk of the target individual, selecting initial spreaders based on the existing centrality metrics may fail to maximize spreading in complex networks.

In this paper, we generalize the traditional problem about the identification of influential spreaders by considering the risk in initial node activation. We assume that the probability to activate nodes acted as initial spreaders decrease with the degree of nodes. For simplicity, we use the exponential decay function to approximate the relation between the activation probability and the degree of nodes. The expected value of outbreak size over the activated probability could quantify the effective spreading coverage of nodes (i.e. the expected value of the number of infected nodes in a spreading initiated from a single node). On random networks, we analyse the  degree value of the optimal initial spreaders by the bond percolation model. The finding suggests that the optimal seed policy (i.e. selecting the optimal initial spreaders to maximize the effective spreading) is not the largest-degree node in the network. Meanwhile, we verify that existing centrality is correlated with the degree on the real-world network. Simply discounting for a degree in existing centralities might not be enough to identify the effective spreaders. Therefore, it is necessary to devise a new method for this case. We then propose a risk-aware method to identify the effective spreader, further maximizing the effective spreading. The performance of the risk-aware centrality is tested on disparate real networks by SIR model. Numerical simulations show that our method outperforms existing benchmark centralities (i.e. the ratio between existing centralities and degree).

\section{Method}

\subsection{\textbf{Maximizing spreading with risk in node activation}} We briefly describe the problem of spreading with the risk in node activation. The basic idea is that it is difficult to convince an individual with more followers in social networks (i.e. larger-degree node) than an individual with fewer followers (i.e. smaller-degree node) to act as the initial spreader. We denote this as the risk of activating the node for spreading. Although the node with a larger degree could disseminate a large fraction of the population, it may refuse to initiate contagion due to the higher activation risk. A natural problem is which node should be selected as the initial spreader for maximizing the spreading under the risk in node activation. 

\begin{figure}[h]
	\centering
	\includegraphics[width =1.0 \textwidth]{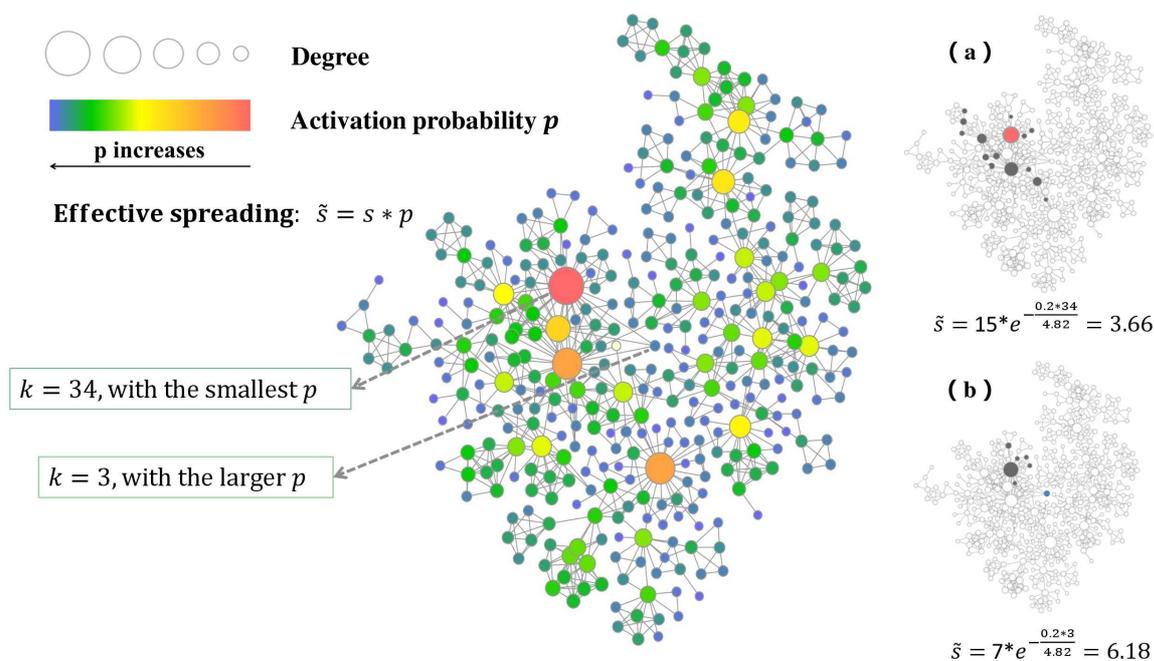}
  	\caption{\textbf{Illustration of the problem for maximizing spreading with risk in node activation.} Here, we assume that the probability in initial node activation decreases with the degree of nodes when one considers the realistic factors (e.g. the cost, accessibility of node) into the problem. The effective spreading coverage could be quantified by the expected value of outbreak size over activation probability. The degree of selected two nodes (red and blue) in the ca-Netscience network are $34$ and $3$ respectively. Although the red node acted as an initial spreader could infect a large number of nodes than the blue one, it has the smallest activation probability in the network. In the subplots (a) and (b), we show the contagion triggered respectively from the two nodes with a critical infection rate by SIR model. Obviously, the number of infected nodes originated from the red node is larger than the blue, but the effective spreading coverage of the red is lower than the blue node ($\lambda = 0.2$).}
\label{figure1}
\end{figure}

To address this problem, we first quantify the activation risk of nodes. Based on the assumption that the activation risk decreases with the degree of nodes, we employ the exponential decay function to characterize the negative relation between degree and activation probability, because its analytical result is tractable and agrees well with the intuition in reality.  A detailed discussion about the selection of the function form could see the \hyperref[sec:Appendix F]{Appendix F}. The exponential decay is monotonic and could map the value of degree (i.e. the number of node's immediate neighbors) into the range from 0 to 1. The activation probability of nodes is acquired by the formula,

\begin{equation}
p_{i}=e^{^{-\frac{\lambda k_{i}}{< k >}}},
\label{1}
\end{equation}
where $k_i$, $\langle k \rangle$, and $\lambda$ are respectively the degree of node $i$, the average degree of network, and the exponential decay constant that is called the risk parameter for determining the difference of activation probability among nodes with distinct degrees. When $\lambda=0$, the problem degenerates to the original definition because each node in the network has the same activation probability. Only if $\lambda>0$, activation risk emerges. A larger $\lambda$ corresponds to a higher risk when trying to activate large-degree nodes as initial spreaders. $p_i$ represents the probability that the node $i$ accepts to act as the initial spreader. Then, the spreading coverage $s_{i}$ is defined as the ratio of infected nodes to all nodes in the network, given the spreading originated from node $i$. When one takes into account the activation risk of initial spreaders, the expected value of  spreading coverage of nodes is naturally regarded as the effective spreading coverage. Thus, the effective spreading coverage of node $i$ can be easily expressed as $\tilde{s_{i}}=p_{i}*s_{i}$.  In this paper, we employ effective spreading coverage as a target function to quantify the practical outbreak size of spreading initiated from a target node with activation risk. The illustration of the problem could be seen in Fig. \ref{figure1}.

\captionsetup[figure]{labelfont={bf},labelformat={default},labelsep=period,name={Fig.}}

\subsection{\textbf{SIR model}} The risk of node activation is more common in real problems such as advertising. We thus describe the spreading process as the information spreading on social networks. Susceptible-infected-recovered (SIR) could well describe the information spreading process in social media \cite{daley1964epidemics, pastor2015epidemic}. Therefore, we apply the SIR model to simulate this process on disparate empirical networks. In the SIR model, individuals are classed into three states: susceptible (S), infected (I), and recovered (R). S nodes do not carry the disease and can be infected. I nodes carry the disease and can infect others. R nodes either die or recover and immune to further infection. At the beginning of dynamics, all nodes are initially in the susceptible state except for an initial infected spreader. At each time step, nodes in the infected state can infect their neighbors with probability $\beta$ who are in the susceptible state, then immediately transform from infected to recover state. The nodes in the recovered state never change their state. The dynamics process continues until there are no infected nodes in the network. At the end of dynamics, the total number of infected nodes is calculated by counting the number of nodes in the recovered state. The spreading coverage of nodes is calculated by the ratio of infected nodes to all nodes in the network. Due to the stochastic nature of the model, all experimental results are obtained by averaging over 1000 independent numerical simulations with the same initial condition. 

\subsection{\textbf{The degree of the optimal initial spreaders to maximize the effective spreading coverage on random network}} The relation between the SIR and the bond percolation model is studied by Newman\cite{newman2002spread}. The SIR model with infection rate $\beta$ is equivalent to a bond percolation model with bond occupation probability $T$. The bond percolation model could give the exact mean size of the SIR epidemic outbreak triggered from a randomly chosen single node by generating function. Here, we investigate the maximum effective spreading coverage on a random network with a given degree distribution by the bond percolation model. Then, we also give the degree of nodes selected as the optimal initial spreader. To this end, we first derive the mean size of the outbreak originated from the single node with degree $k$. The formula could be seen as following,
\begin{equation}
	\langle s_{k} \rangle =1+\frac{k\beta}{1-\beta G_{1}^{'}(1)},G_{1}^{'}(1)=\frac{\langle k^2\rangle-\langle k\rangle }{\langle k\rangle}, if\quad\beta<\beta_{c},
	\label{2}
\end{equation}
where $\langle k\rangle$, $\langle k^2\rangle$, and  $\beta_c$ are the average degree of network, second moment of degree, critical infection rate respectively, and $\beta_c=\frac{\langle k\rangle}{\langle k^2\rangle-\langle k\rangle}$. $G_1(x)$ denotes the generation function of the degree distribution of nodes reached by following a randomly chosen edge, $G_1(x) = \sum_{k=1}^{\infty}\frac{kp_k}{<k>}x^{k-1}$. $G_1^{'}(1)$ means the derivative of $G_1(x)$ at $x =  1$. The detailed description about $G_1(x)$ and the derivation of Eqs. \ref{2} see the \hyperref[sec: Appendix A]{Appendix A}. Here, we only consider the part where the infection rate is less than the critical infection rate ($\beta<\beta_{c}$). In the case of larger $\beta$,  the role of individual nodes is no longer important since the final spreading coverage is independent of where it originated from. Combining the Eqs. \ref{1} with \ref{2}, we have
\begin{equation}
	\langle \tilde{s_{k}}\rangle =e^{-\frac{\lambda k}{\langle k\rangle}}[1+\frac{k\beta}{1-\beta G_{1}^{'}(1)}].
	\label{3}
\end{equation}   
Eqs. \ref{3} represents the mean size of effective spreading coverage initiated from a single node with degree $k$ on a random network. By differentiating for $k$, we have
\begin{equation}
	\frac{\partial \langle\tilde{s_k}\rangle}{\partial k} = e^{\frac{\lambda k }{\langle k\rangle}}[-\frac{\lambda}{\langle k \rangle}(1+\frac{k\beta}{1-\beta G_{1}^{'}(1)})+\frac{\beta}{1-\beta G_{1}^{'}(1)}].
	\label{4}
\end{equation}
The degree of optimal node ($k^*$) corresponding to the maximum effective spreading coverage $\langle \tilde{s}^* \rangle$ can be obtained by setting the $\frac{\partial \langle\tilde{s_k}\rangle}{\partial x}=0$,
\begin{equation}
	k^*=\frac{\langle k\rangle}{\lambda}-\frac{1-\beta G_{1}^{'}(1)}{\beta}.
	\label{5}
\end{equation}
Substituting the $G_{1}^{'} (1)=\frac{1}{\beta_c}$  into the Eqs.\ref{5}, we further simplify the equation,
\begin{equation}
	k^{*}=\frac{\langle k \rangle}{\lambda}-\frac{1-\frac{\beta}{\beta_{c}}}{\beta}.
	\label{6}
\end{equation}
We find that the degree of optimal nodes depends on the infection rate $\beta$ and the risk parameter $\lambda$, because $\langle k\rangle$ and $\beta_c$ is a constant with a given degree distribution. The result suggests that the optimal initial spreaders to maximize the effective spreading coverage is different from the traditional problem (i.e. influence maximization problem on random network). Besides, substituting the Eqs.\ref{6} into Eqs.\ref{3}, we also give the mean size of maximum effective spreading coverage $\tilde{s}^*$ triggered from nodes with the $k^*$ on a random network with a given degree distribution by
\begin{equation}
	\langle \tilde{s}^* \rangle = e^{-[1-\frac{\lambda}{\beta \langle k\rangle}+\frac{\lambda}{\beta_c \langle k\rangle}]}\frac{\beta_c \beta \langle k \rangle}{\lambda(\beta_{c}-\beta)}
\end{equation}

We further analyze $k^*$. In the Eqs.\ref{6}, $k^*$ is codetermined by $\lambda$ and $\beta$. Specifically, we wish to analyze the effect of variable on the degree of optimal spreaders by fixing a parameter. (1) If we take the limit of $\beta$ as $\lim_{\beta \to \beta_{c}} \frac{\langle k \rangle}{\lambda}-\frac{1-\frac{\beta}{\beta_c}}{\beta} $, the optimal degree $k^* = \frac{\langle k \rangle}{\lambda}$. When $\lambda\to 0$, $k^*=\infty$. When $\lambda\to \infty$, then $k^*=0$. The optimal degree of nodes decrease with $\lambda$ by fixing $\beta$. This tells us that in a higher-risk condition, the optimal spreaders shift to select the lower-degree nodes as the initial spreader. (2) By setting the $\lambda =\langle k \rangle$, the optimal degree $k^*=\frac{1}{\beta_c} -\frac{1}{\beta}$. One could see that $k^{*}$ increases monotonically with $\beta$, suggesting that as the infection rate increases, the larger-degree nodes have a comparative advantage for maximizing the effective spreading coverage.

\subsection{\textbf{Risk-aware metrics}}
In the related work, some centrality metrics are proposed to identify influential spreaders or important nodes\cite{chen2012identifying,kitsak2010identification,liao2017ranking,zhou2019fast}. When the risk is considered in node activation, the analysis of the optimal initial spreaders on random networks reveals the fact that optimal seed policy is obviously distinct from the largest degree node in the original problem, which suggests that it might be not efficient to directly utilize existing metrics to identify the initial spreaders in the real network. Meanwhile, the correlation analysis between degree and other metrics is made to verify that existing centrality metrics are not enough to perform well in the identification of effective spreaders because they are correlated with the degree (for correlation analysis, see the \hyperref[sec: Appendix C]{Appendix C}). Therefore, it is necessary for maximizing the effective spreading to design a new measure in real networks. 

The analytic result shows that the degree value of the optimal  spreaders in the problem is inversely proportional to risk parameter $\lambda$ and has a positive relation with infection rate $\beta$, suggesting small-degree nodes linked to many hubs are more likely to maximize the effective spreading. Inspired by the analytic result, we consider two factors to design the new metric, i.e. outbreak size and activation risk. On one hand, we expect that it could select nodes linked to many hubs, because the spreading initiated from those nodes could cover more nodes with the help of the neighboring hub when the infection rate $\beta$ is larger, which also requires that degree of the node itself could not be too small (This corresponds to the analytical solution where the optimal degree value has a positive relation with the infection rate $\beta$). On the other hand, we hope that the initial spreaders are selected as small-degree nodes when the activation risk $\lambda$ is higher (This corresponds to the analytical solution where the optimal degree value is inversely proportional to risk parameter $\lambda$). Therefore, the effective spreaders could be better characterized by the following two aspects: (1) spreaders with a strong spreading ability (e.g. possessing many high influence neighbors), (2) spreaders associated with lower activation risk (e.g. smaller degree nodes). 

In this paper, we propose the risk-aware metric ($RA$) to identify the efficient spreaders by rewarding nodes with higher-degree neighbors and penalizing higher-degree nodes. The risk-aware metric of node $i$ is defined as follow,
\begin{equation}
	RA_{i}=\sum_{j\in \tau(i)}{(\frac{k_{j}}{k_i+k_j})^{\theta}},
\end{equation}
where $k_i$, $\tau(i)$, $k_j$ are the degree of node $i$, the neighbor set of node $i$, and the degree of node $j$ respectively. $\frac{k_j}{k_i+k_j}$ means the potential influence of node $i$ obtained from neighbor $j$. The potential influence means that the node's neighbors have higher influence to initiate a spreading although the node itself has lower influence. If $k_i=k_j$, $\frac{k_j}{k_i+k_j}=1/2$. If $k_i\ll k_j, \frac{k_j}{k_i+k_j}\approx 1$. If $k_i \gg  k_j$, then $\frac{k_j}{k_i+k_j}\approx 0$. As a consequence, $\frac{k_j}{k_i+k_j}\in (0,1)$. In this paper, the risk parameter $\lambda$ to determine the risk difference of nodes with distinct degree is incorporated into the defined problem. To identify the effective spreader under different conditions (i.e. various $\lambda$), accordingly, the parameter $\theta$ is introduced to adjust the potential influence obtained from different neighbors. When $\theta=0$, the potential influence obtained from different neighbors is equal, and the risk-aware metric degenerates to the degree. When $\theta > 0$, the lower potential influence obtained from neighbors will be largely weakened with an increase of $\theta$. In other words, small-degree nodes are likely to have larger potential influence, because this term $(\frac{k_j}{k_i+k_j})^{\theta}$ plays an important role in the contribution of $RA$ than the number of neighbors.
Intuitively, a node has a large value of $RA$ if it is connected to many other nodes that have a higher degree than the node. In fact, it is hard to estimate which nodes have the large value of $RA$. For instance, when the degree of node $i$ is rather small, the potential influence obtained from each neighbor is relatively large. On the other hand, the overall sum of potential influence from neighbors is small due to the few neighbors.

\begin{figure}[h]
	\includegraphics[width =1.0 \textwidth]{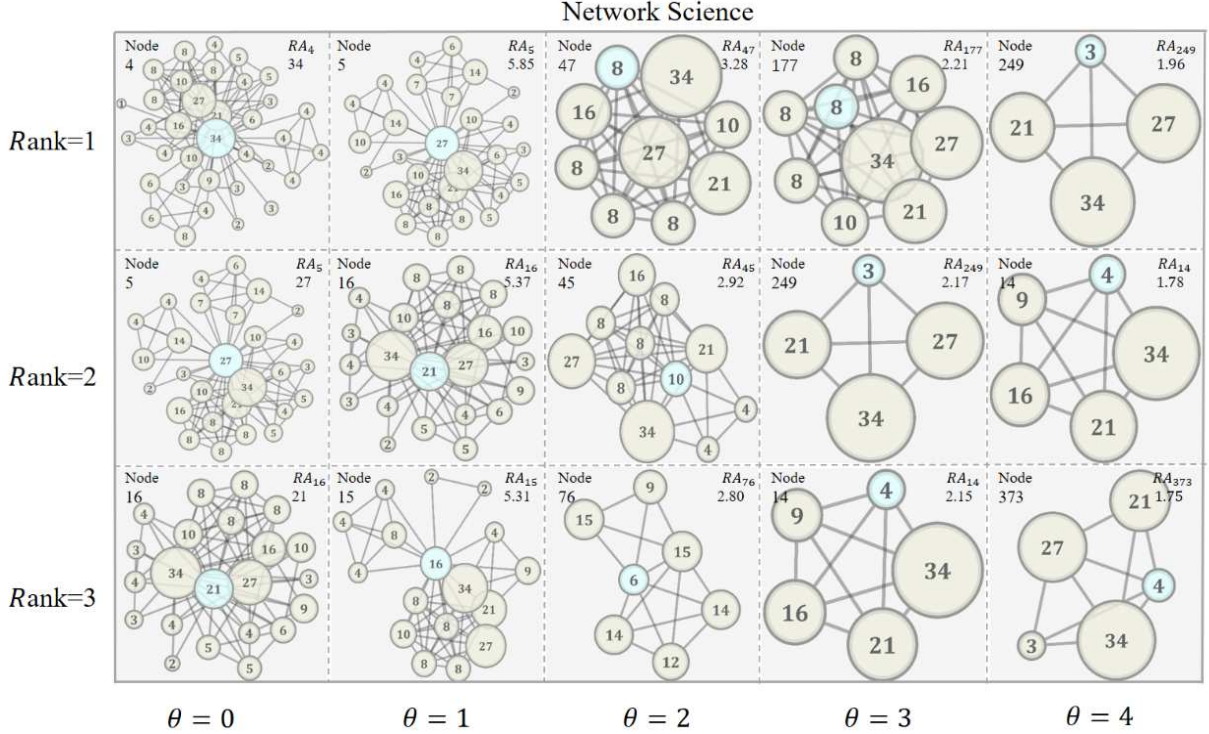}
\caption{\textbf{Illustration of top-3 nodes ranked by the risk-aware metric for different parameter $\theta$ in ca-Netscience network}. In each inset, we display the identified target node (blue node) and its neighbors (yellow node) extracted from the original network as well as the link between them. The number in node denotes degree in the original network and the size of node shape also denotes the degree (note that the links of neighbor nodes do not show entirely in each inset).  The number on the upper left and upper right corner respectively suggest the node label and $RA$ score. By observing the local structure of each identified node, it is easier to understand the concept of the risk-aware metric.}
\label{figure2}
\end{figure}

To further understand the metric, we make an illustration of nodes ranked by the risk-aware metric for different parameters $\theta$ in Fig. \ref{figure2}. One could see clearly how the most highly ranked node identified by $RA$ varies with $\theta$. As $\theta$ increases, top ranked nodes are likely to be small-degree nodes with higher-degree neighbors. Actually, $\theta$ determines the ability to identify the potential influence of nodes. The larger the value of $\theta$ is, the greater the potential influence of the identified node is. The $RA$ is significantly different from the traditional centrality metrics that measure the importance of nodes. The computational complexity to traverse the neighbors of a node is simply the average degree $\langle k \rangle$ of networks. If one estimates the potential influence of each node in a network by considering the degree of node and their neighbors, the computation complexity is $O(N {\langle k\rangle}^2)$ where $N$ is the number of nodes in the network.

Based on the idea to reward nodes connected to higher-degree nodes and penalize the nodes with high degree, we further consider other metrics to identify the effective spreader by different function form. The first potential influence metric ($PI\_1$) and second potential influence metric ($PI\_2$) are respectively defined in Eqs.(9) and Eqs.(10),
\begin{equation}
	PI\_1=\sum_{j\in \tau(i)}{k_j-k_i},
\end{equation}
\begin{equation}
	PI\_2=\sum_{j\in \tau(i)}{e^{k_j-k_i}},
\end{equation}
where $k_i$, $\tau(i)$, $k_j$ are the degree of node $i$, the neighbor set of node $i$, and the degree of node $j$ respectively. Two potential influence metrics are parameter-free. $PI\_1 \in (-\infty,+\infty)$, $PI\_2 \in (0,+\infty)$. Although $PI\_1$ might be negative, we still could rank nodes.

\subsection{\textbf{Centrality metrics}} 
To confirm the effectiveness of new metrics, we need to compare them with some baseline methods. Due to the fact that there is a higher correlation between existing centrality metrics and degree (for correlation analysis, see \hyperref[sec: Appendix C]{Appendix C}), it is not fair to directly utilize the existing metrics as baseline methods to compare with new metrics. Here, we calculate the ratio between existing centrality metrics and degree as baseline methods (e.g. Katz score / degree). In this way, it is reasonable to compare new metrics with baseline methods, which could be proved in the correlation analysis (see the Fig.\ref{Sfigure1} in \hyperref[sec: Appendix C]{Appendix C}). In this paper, we employ some representative centrality metrics to obtain the baseline method. We briefly introduce these metrics.
	
\textbf{(1) Degree}. The degree of node $i$ is defined as the number of immediate neighbors. $k(i) = \sum_{j}^{N} a_{ij}$ where $a_{ij}$ is a component of the network's adjacency matrix and $N$ is the number of nodes in the network. The degree reflects the direct influence of this node.

\textbf{(2) Coreness (also called k-shell, $KS$)}\cite{kitsak2010identification}. The k-shell reflects the location of a node, which is regarded as more important than the degree in evaluating its spreading influence. If a node is located in the core part of the network, the influence of the node will be higher than the node that is located in the periphery. Nodes are assigned to k-shell by the following steps: (1) In the first step, all nodes with degree $k=1$ are removed, which will cause the reduction of degree value for remaining nodes. The process will continue by successive pruning of nodes until no more nodes with degree $k=1$ remain, and then assign all removed nodes to the 1-shell. In this process, for each remaining node, residual degree refers to the number of links connecting to other remaining nodes, and exhausted degree denotes the number of links connecting to removed nodes. (2) In the second step, all remaining nodes with degree $k=2$ will be removed, then the process will be iteratively updated until the residual degree of all remaining nodes is larger than $2$, the k-shell of removed nodes in the second step is equal to $2$. (3) The process of decomposition will continue until all nodes in the network are removed. The k-shell of each node corresponds to its shell layer.

\textbf{(3) Closeness}\cite{sabidussi1966centrality}. The closeness is defined as the inverse of the mean geodesic distances form a node to all other nodes. $CC_i=\frac{1}{N-1}\sum_{j,j\neq i}^{N}{\frac{1}{d_{ij}}}$  where $N$ is the number of nodes in the network and $d_{ij}$ denotes the length of shortest path between node $i$ and node $j$. Obviously, the larger the closeness is, the more central the node is. 

\textbf{(4) Betweenness}\cite{freeman1978centrality}. The betweenness measures the number of times that a node acts as a bridge along the shortest path between the other two nodes. It is defined as $BC_i = \sum_{j,j \neq i}^{N}\sum_{k, k\neq i,k\neq j}^{N}\frac{{\sigma_{jk|i}}}{\sigma_{jk}}$ where $\sigma_{jk}$ denotes the number of shortest paths between node $j$ and node $k$ and $\sigma_{jk|i}$ denotes the number of path which pass through node $i$ among all shortest paths between node $j$ and node $k$. $N$ is the number of nodes in the network.

\textbf{(5) Eigenvector}\cite{bonacich2007some}. The eigenvector centrality considers not only the number of immediate neighbors but also the influence of each neighbor. The eigenvector centrality of node $i$ is denoted by $x_i$, $x_i = \frac{1}{\lambda} \sum_{j=1}^{N}a_{ij}x_j$ where $\lambda$ and $a_{ij}$ is respectively the largest eigenvalue and a component of adjacency matrix $A$, and $x_j$ denotes the eigenvector centrality of node $j$, which could be described by the matrix form as $\overrightarrow{x} =\frac{1}{\lambda} A \overrightarrow{x}$. The eigenvector is widely used to measure the influence of nodes.  

\textbf{(6) Subgraph}\cite{estrada2005subgraph}. The subgraph centrality is defined as a weighted sum of the number of all closed walks starting and ending at the node $i$. The subgraph centrality of node $i$ is defined as  $SC_i = \sum_{k}^{\infty}\frac{(A^k)_{ii}}{k!}$
	where $k$ is the length of closed walks. $(A^k)_{ii}$ represents the number of closed walks with length $k$ starting and ending at node $i$, which is defined as the $i$ the diagonal entry of $k$ the power of adjacency matrix $A$. The Closed walks with shorter lengths have more influence on the centrality than longer closed walks in the subgraph centrality.

\textbf{(7) Katz}\cite{zhan2017identification}. The Katz centrality of nodes is defined by considering all different length walks. It holds that a shorter walk plays a more important role than a long walk. The specific formula is $KC_i = \sum_{j}^{N}\sum_{k}^{\infty}s^k(A^k)_{ij}$ where $s \in (0,1)$ is a tunable parameter and  $s^k$ denotes the weight of a walk with length $k$. For the power method to converge, the value of the attenuation factor $s$ has to be set such that it is smaller than the reciprocal of the absolute value of the largest eigenvalue of $A$.

\textbf{(8) Collective influence ($CI$)} \cite{morone2015influence}.  Collective influence pursues the goal of maximizing the overall influence of multiple spreaders. It could give a minimal set of spreaders for the objective. The $CI$ is an adaptive algorithms which removes the nodes progressively according to current $CI$ value, defined as $CI_{\ell}(i)=(k_i-1)\sum_{j\in \partial B(i,\ell)}(k_j-1)$ where $k_i$ is the degree of node $i$, $B(i,\ell)$ is the ball of radius $\ell$ centered on node $i$, and $\partial B(i,\ell)$ is the frontier of the ball, that is, the set of nodes with shortest path length $\ell$ from the node $i$. The  $\ell$ is a non-negative integer which does not exceed the network diameter. 

\textbf{(9) Non-backtracking ($NB$)}
\cite{PhysRevE.93.062314, PhysRevE.90.052808}. The nonbacktracking centrality mainly considers the local effect of eigenvector centrality, because a hub node with a higher eigenvector centrality distributes the centrality to its neighbors, and then back it again and inflate the hub's centrality. The nonbacktracking centrality prevents the reflection and excludes self effect in summation over neighbors. It can be calculated by the nonbacktracking matrix $B$ (the nonbacktracking matric is converted by the adjacency matrix $A$, see the reference \cite{Krzakala20935,PhysRevE.90.052808}). The nonbacktracking centrality $x_j$ of node $j$ is defined to be the sum of centralities over the neighbors of $j$, $x_j = \sum_{i}A_{ij}v_{i \rightarrow j}$ where the $v_{i \rightarrow j}$ is the element of leading eigenvector of the nonbacktracking matrix $B$ and give the centrality of node $i$ ignoring any contribution from $j$.

In the simulation experiment, the degree, $ks$, closeness, betweenness, eigenvector, subgraph, and Katz centrality are implemented by the networkx package in Python. We implement the $CI$ and $NB$ centrality based on the original definition of metrics. In the collective influence centrality, we set the radius $\ell=3$, because $\ell$ needs to be set smaller than the diameter of the network (the smallest diameter among disparate networks in this paper is 5, see the Table \ref{table1} in \hyperref[sec: Network]{Appendix B}).

\subsection{\textbf{Evaluation metrics}}
To examine the performance of methods, we employ the Kendall rank correlation coefficient to estimate the ability of centrality metrics to identify the effective spreaders. In addition, we also use the average of effective spreading coverage to test the performance of methods for identifying top n effective spreaders. To obtain a comprehensive understanding of method performance, we devise the normalized score to summarize the performance of each method in all networks employed in this paper. 

(1) \textbf{Kendall rank correlation coefficient ($\tau$)} \cite{kendall1938new}. It is also called as Kendall's $\tau$ coefficient used to assess statistical associations based on the ranks of the data. The $\tau$ test is a non-parametric hypothesis test for statistical dependence based on the $\tau$ coefficient. For any pair of observations $(x_i,y_i)$ and $(x_j, y_j)$, they are said to be concordant pairs if $(x_j-x_i)(y_j-y_i) > 0, \forall j>i,i,j=1,2,...,n$. They are said to be discordant pairs if $(x_j-x_i)(y_j-y_i) < 0, \forall j>i,i,j=1,2,...,n$. The Kendall's $\tau$ coefficient is defined as $\tau = \frac{N_c-N_d}{\tbinom{n}{2}} = \frac{2(N_c-N_d)}{n(n-1)}$ where $N_c$ is the number of concordant pairs and $N_d$ is the number of discordant pairs. Here, we use the Kendall rank correlation to investigate how the ranking based on centralities is correlated to the ranking generated by effective spreading coverage. According to the definition of Kendall rank correlation, $-1 \leq  \tau \leq  1$. The Kendall's $\tau$ coefficient will be 1 if the agreement between centralities and effective spreading is perfect, which means that the employed centrality metrics could well identify the effective spreader. 

(2) \textbf{Average of effective spreading coverage ($\langle \tilde{s_n} \rangle$).}
To systematically estimate the performance of method to identify the effective spreaders under various risk, we average the 
effective spreading coverage $\langle \tilde{s} \rangle$ over various $\lambda$ ($\lambda \in [0,0.9]$ with a step of 0.1). When dealing with actual problems, we might only need to identify high-ranking effective spreaders rather than all. To estimate the performance of method in identifying high-ranking effective spreaders, we thus average the effective spreading coverage over top n effective spreaders. The formula could be seen in the Eqs.\ref{eqs10},
\begin{equation}
	\label{eqs10}
	\langle \tilde s_n \rangle= \frac{1}{|\sigma(top_n)|}\sum_{i\in\sigma(top_n)}\sum_{\lambda=0}^{\lambda=0.9}{\frac{\tilde{s_i}(\lambda)}{10}},
\end{equation}
where $\sigma(top_n)$ denotes the set of nodes whose centrality score is ranked in top n (e.g. n=1,10,20), $\tilde{s_i}(\lambda)$ means the effective spreading coverage when risk parameter is set as $\lambda$.

(3) \textbf{Normalized score ($NS$).} It is designed to summarize the performance of metrics in all networks, further having a comprehensive understanding for different methods about overall performance. We normalize the performance of metrics in each networks such that all metrics's performance range in $[0,1]$, and then average the normalized performance across networks. The normalized score is defined in the Eqs.\ref{eqs100},
\begin{equation}
	\label{eqs100}
	 NS_m(e) = \frac{\sum_{i \in \gamma(n)}\frac{c_m^i(e)-c_{min}^i(e)}{c_{max}^i(e)-c_{min}^i(e)}}{|\gamma(n)|}, m \in \gamma(c),
\end{equation}
where $e$ represents the evaluation metrics (e.g. Kendall's $\tau$,$\langle \tilde{s_n} \rangle$), $\gamma(n)$ and $\gamma(c)$ are respectively the set of networks and centrality metrics. $c_{min}^i(e)$ denotes the value of the worst-performing metrics among all centralities in $i$ network according to the $e$ evaluation metrics. Inversely, $c_{max}^i(e)$ denotes the value of the best-performing metrics among all centralities in $i$ network according to the $e$ evaluation metrics. Similarly, $c_{m}^i(e)$ is the value of the $m$ metric in $i$ network according to the $e$ evaluation metrics. $|\gamma(n)|$ is the number of networks in the datasets. Based on $NS_m(e)$, we could give a normalized score for different methods according to a given estimation metrics, which quantifies overall performance in all networks. $NS_m(e) \in [0,1]$. For Kendall's $\tau$ and $\langle \tilde{s_n} \rangle$, the larger the normalized score is, the better centrality metrics perform in all networks.

\section{Dataset}
To confirm the correctness of analytic result on random networks and validate the risk-aware method, we respectively conduct experiments on the Erdos-Renyi network (ER) and 40 real networks. The size of these networks range from 34 to 4991, and their average degree varies between 2.04 and 82.42. We consider 40 networks of small to medium size from different systems. Specifically, these networks include  6 biology networks (bio-Yeast\cite{jeong2001lethality}, bio-Celegans, bio-Grid-Plant, bio-Grid-Worm, Protein\cite{Yu104}, Metabolic\cite{schellenberger2010bigg}), 4 collaboration networks (ca-Erdos992, ca-GrQc, ca-Netscience\cite{newman2006finding}, ca-CSphd), 4 animal networks (ani-Dolphins\cite{konect:dolphins}, ani-Mammalia, ani-Aves-Songbird, ani-Reptilia), 4 email networks (email-Dnc, email-Corecipient, email-Enron-Only, email-Univ), 4 infrastructure networks (inf-Power\cite{watts1998collective},inf-Euroroad\cite{bader2012graph},inf-Usair97, inf-Openflights), 4 facebook networks (socfb-Calrech36, socfb-Haverford76, socfb-Reed98, socfb-Simmons81), 3 ecology networks (econ-Mahindas, econ-Wml, econ-Poli), 3 human social networks (hs-Arenas-Jazz\cite{konect:arenas-jazz}, hs-Physical\cite{konect:coleman1957}, hs-Zachary\cite{konect:ucidata-zachary}), 3 interaction networks (ia-Crime-Moreno, ia-Fb-Messages, ia-Infect-Dublin), 2 retweet networks (rt-Twitter-Copen, rt-Retweet), 1 brain network (bn-Mouse-Kasthuri), 1 web network (web-EPA), and 1 social network (soc-Karate). The networks with unlabeled reference are download from the network repository\cite{nr}. Details about analyzed networks can be found in \hyperref[sec: Network]{Appendix B}.
	
\section{Result}

\subsection{\textbf{The degree of the optimal initial spreaders on ER network}} 
\begin{figure}[!h]
	\centering
	\includegraphics[width =1.0 \textwidth]{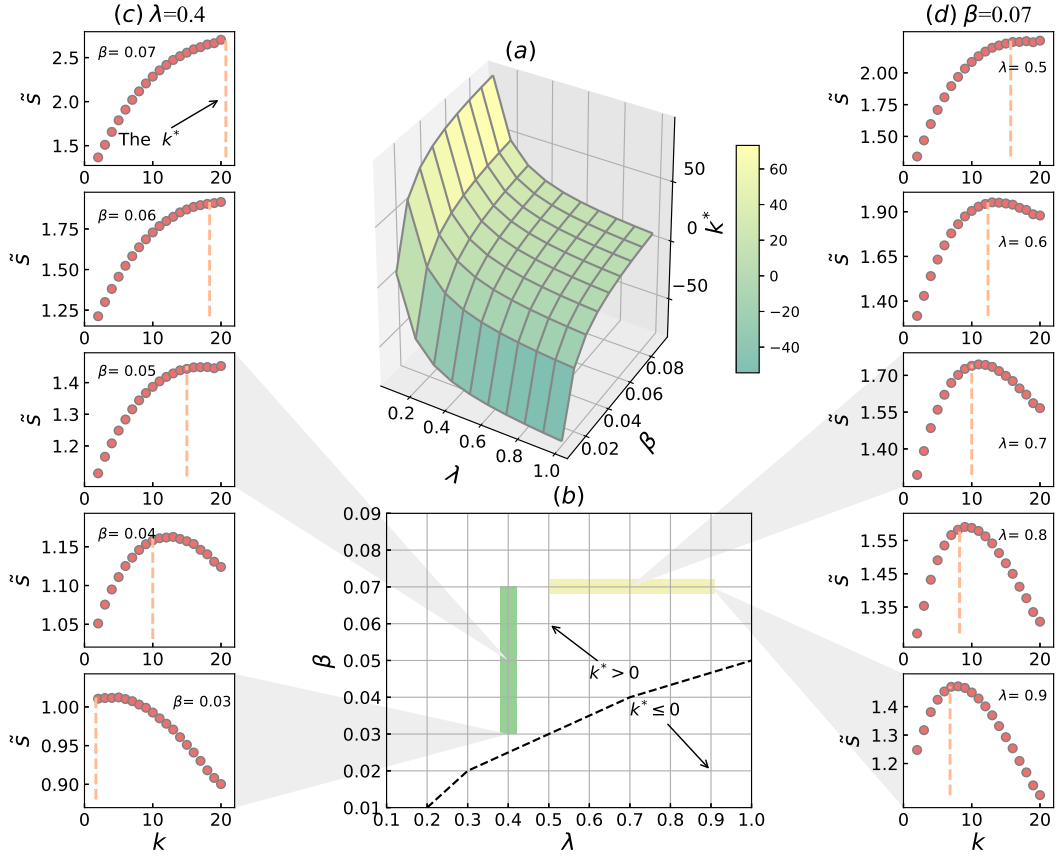}
	\caption{\textbf{The degree of the optimal initial spreaders to maximize the effective spreading coverage on ER networks}. We set the parameters of ER network with $N = 1000$ and $p_c = 0.01$. The mean degree of ER network is $\langle k\rangle = Np_c = 10$. $\left(\textbf{a}\right)$ The optimal degree $k^*$ plotted as a function of risk parameter $\lambda$ and infection rate $\beta$. $\left(\textbf{b}\right)$ The parameter space of $\lambda$ and $\beta$, the dashed line marks the region where the $k^*$ greater than 0.  $\left(\textbf{c}\right)$$\left(\textbf{d}\right)$ The effective spreading obtained from the numerical simulation of SIR model as a function of degree  $k$ on ER networks. The point in each subfigure is the average of 500 ER networks. The orange dashed line denotes the exact $k^*$ calculated by the Eqs.\ref{9}}
	\label{figure3}
\end{figure}

Taking the ER network as an example, we verify the degree of the optimal spreaders to the maximum effective spreading coverage. The degree distribution of the ER network is Poisson for larger $N$ where $N$ is the total number of nodes in the network. The mean degree of the ER network is $\langle k \rangle = Np_c$ where $p_c$ refers to the probability of edge creation. The critical infection rate is $\beta_{c} =\frac{<k>}{<k^2>-<k>}= \frac{1}{Np_c}$ because $\langle k^2 \rangle = \langle k \rangle + {\langle k \rangle}^2$ on ER network. Substituting the $\beta_{c}$ and $\langle k \rangle$ into the Eqs.\ref{6}, we obtain the equation
\begin{equation}
		k^{*}=Np_c+\frac{Np_c}{\lambda}-\frac{1}{\beta}.
	\label{9}
\end{equation}
Given an ER network with $N$ and $p_c$, we find that $k^{*}$ is determined by $\lambda$ and $\beta$. In Fig. \ref{figure3} $\left(a\right)$, we show the optimal degree plotted as a function of $\lambda$ and $\beta$. One could see clearly that $k^*$ is inversely proportional to $\lambda$ and has a positive relation with $\beta$. The result reveals an important finding that there is an optimal initial spreader to maximize the effective spreading for any pair of parameters $\lambda$ and $\beta$. Once the risk is considered into the problem, the most effective spreader does not correspond to the largest-degree node in networks but be determined by $\lambda$ and $\beta$. Meanwhile, we also marked the parameter region where $k^* > 0$ in Fig.\ref{figure3} $\left(b\right)$, since the degree value of nodes in the connected network is greater than 0. When we take the pair of parameters located in the region where $k^* \leq 0$, the optimal degree value of nodes would be 1. 

In Fig \ref{figure3} $\left(c\right)$ and $\left(d\right)$, we show the effective spreading coverage $\tilde s$ for several pair of parameters, which is obtained from the exact numerical simulation of SIR model on ER networks (see the green and yellow stripe in the Fig.\ref{3} $\left(b\right)$). Simulations are performed on the network of $N$ = 1000 with $p_c = 0.01$. For each pair of parameter $\lambda$ and $\beta$, we simulate the spreading triggered from a single node in the network and conduct 1000 experiments for each node. Then we calculate the effective spreading coverage $\tilde s$ by averaging over nodes with the same degree. The simulation results show two important findings analyzed in the previous section. First, $k^*$ corresponding to the maximum effective spreading in the data agrees well with our analytic results (the orange dashed line). Also, the changing trend of $k^*$ in the simulation for different pairs of parameters consistent with the analytic result, confirming the correctness of degree of the optimal initial spreaders. The small disagreement between simulations and analytic results for $k^*$ appears to be a finite size effect due to the relatively small network size in the simulation. Second, one could see that $k^*$ increases with $\beta$ when fixing $\lambda$, suggesting that it is still a better choice to target a high-degree node as initial spreaders if the infection rate is larger even if there is risk in node activation. On the contrary, with a given $\beta$, $k^*$ decreases with $\lambda$, indicating that one should choose a conservative seed policy when the risk difference for nodes with various degrees is rather larger, e.g. selecting a small-degree node as an initial spreader.  

\subsection{\textbf{Performance of the risk-aware metric on real networks}} 
\begin{figure}[!h]
	\centering
	\includegraphics[width =1.0 \textwidth]{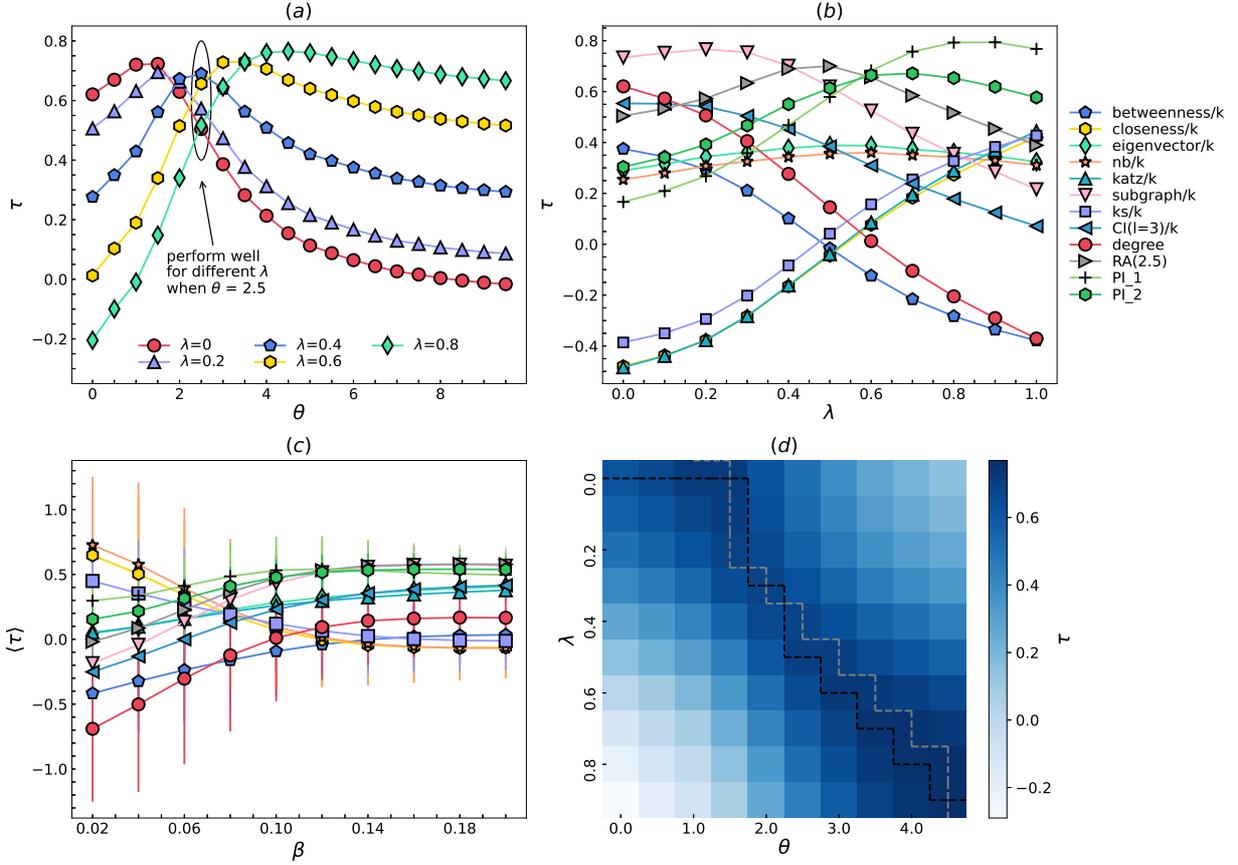}
	\caption{\textbf{The Kendall rank correlation ($\tau$) between different methods and the effective spreading coverage $\tilde s$ on the ca-Netscience network.} $\left(\textbf{a}\right)$ The $\tau$ of risk-aware metric for various $\lambda$ under distinct $\theta$. $\left(\textbf{b}\right)$ The $\tau$ of different methods plotted as a function of $\lambda$. $\left(\textbf{c}\right)$ The $\langle \tau\rangle$ of different methods plotted as a function of $\beta$. The error bar is standard deviation.  $\left(\textbf{d}\right)$ The $\tau$ plotted as a function of $\lambda$ and $\theta$. In $\left(\textbf{a}\right)$ $\left(\textbf{b}\right)$ and $\left(\textbf{d}\right)$, the results are obtained with a critical infection rate $\beta_{c}$.}
	\label{figure4}
\end{figure}

For different methods, we could generate the final ranking of nodes. In principle, the ranking generated by a benchmark centrality metric with good performance should be as consistent as possible with the ranking based on the node's effective spreading coverage $\tilde{s}$. Here, we use the Kendall rank correlation ($\tau$) to measure the performance of different methods. Due to the parameter $\theta$ incorporated into risk-aware metric, we firstly study how $\theta$ affects the performance of $RA$ under different $\lambda$. In the Fig.\ref{figure4} $\left(a\right)$, we show the $\tau$ value of $RA(\theta)$ for diverse risk parameter $\lambda$ on ca-Netscience network. For any $\lambda$, there is an optimal parameter $\theta^*$ to maximize the $\tau$ under the current resolution of $\theta$ with the steps of 0.5. We find that the $\theta^*$ increases with the $\lambda$, suggesting that $RA$ with larger $\theta$ performs better in this case where the activation risk of nodes with a large degree is far higher than the small degree. This is because potential influential nodes connected to a few high-degree nodes could be given a higher centrality value by $RA$ with larger $\theta$, which maintains a higher consistency order with the effective spreading coverage of nodes when the $\lambda$ is larger. Although the $\theta^*$ varies with $\lambda$ within the range from 0 to 0.9, we could fixe a  parameter to make $RA$ achieve the higher $\tau$ as far as possible under different $\lambda$. When  $\theta$ is set as 2.5, the risk-aware metric performs relatively well under different $\lambda$ because the stand deviation of $RA(\theta=2.5)$ is the lowest $\left(S_{\tau}(\theta) =\sum_{\lambda=0}^{0.9}{\frac{(\tau_{\lambda}(\theta)-\langle \tau(\theta)\rangle)^2}{9}}\right)$. In this way, we further compare the $RA(\theta=2.5$) with other baseline methods under different $\lambda$ in the Fig.\ref{figure4} $\left(b\right)$. The $\tau$ of $RA(\theta=2.5)$ is lower than the Subgraph, Katz, Degree, and K-shell with a smaller $\lambda$, but $RA(\theta=2.5)$ remarkably outperforms others with the increase of $\lambda$, indicating that our method performs more accurate than other methods in ranking the effective spreading of nodes. Besides, the $\tau$ of $PI\_1$ and $PI\_2$ is lower than $RA(\theta=2.5)$ when $\lambda < 0.6$, because they penalize high-degree node stronger than $RA$. This could be used to explain why $\tau$ of $PI\_1$ and $PI\_2$ is higher than $RA$ for a larger $\lambda$. Some statistical significance tests are made to verify the Kendall rank correlation between centralities value and effective spreading of nodes (see the Table \ref{table2} in \hyperref[sec: Appendix D]{Appendix D}). In fact, we cannot exactly determine the value of $\lambda$ for the practical problem. In order to systematically study the performance of $RA(\theta=2.5)$, we calculate the $\langle \tau \rangle$ by averaging all $\tau$ over different $\lambda$ we consider, $\langle \tau \rangle = \sum_{\lambda=0}^{0.9}{\tau(\lambda)}/10$, which could quantify the performance of different methods to identify the effective spreader in distinct conditions. Therefore, we further compare $RA(2.5)$ with other methods under different infection rate  $\beta$ in the Fig.\ref{figure4}  $\left(c\right)$. By visual observation, although the $\langle \tau \rangle$ of $RA(\theta=2.5)$ for a larger $\beta$ is the same as subgraph benchmark centrality, the K-S test confirms that the Kendall rank correlation distribution under various conditions is significantly different (see the Table \ref{table4} in \hyperref[sec: Appendix D]{Appendix D}). The result shows that $RA(\theta=2.5)$ outperforms most baseline methods when the infection rate $\beta$ is larger. Meanwhile, the standard deviation of $\tau$ for $RA(2.5)$ is rather low compared with other methods (see the error bar in Fig.\ref{figure4} $(c)$), further revealing that our method is robust under different conditions. Moreover, our method has a huge advantage over global centrality metrics because the $RA$ associated with every node is computed by considering their nearest neighbors. Finally, we investigate how $\lambda$ and $\theta$ together affect the value of $\tau$ in Fig.\ref{figure4} $\left(d\right)$. The black dashed line shows the optimal $\lambda$ corresponding to the maximum $\tau$ varies with the increase of $\theta$, suggesting that $RA$ have a better performance on larger $\lambda$ when $\theta$ gradually increases. Similarly, the grey dashed line denotes the optimal $\theta$ corresponding to the maximum $\tau$ varies with $\lambda$, meaning that a larger $\theta$ should be selected to identify the effective spreaders when $\lambda$ gradually increases. The two dashed lines together reveal that there is a positive correlation between $\lambda$ and $\theta$. For the higher difference of activation risk, $RA$ with a larger $\theta$ might perform better on identifying the effective spreaders. The test of Kendall rank correlation between $RA$ and effective spreading confirms that the correlation for most pair of parameters ($\theta$ and $\lambda$) is significant and the result could be accepted (see the Table \ref{table3} in \hyperref[sec: Appendix D]{Appendix D}).

\begin{figure}[!h]
	\centering
	\includegraphics[width =1.0 \textwidth]{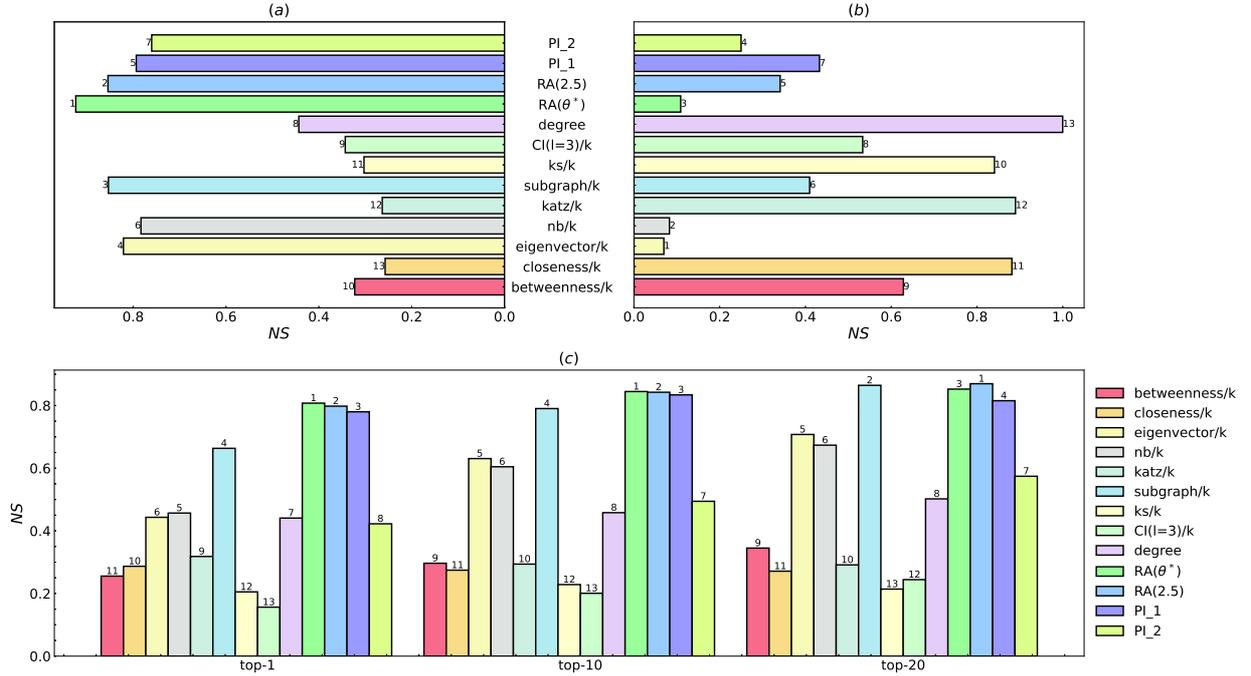}
    \caption{\textbf{The normalized score ($NS$) of evaluation metrics for different methods in 40 networks}. The number on the bar denotes the ranking among all methods. $\left(\textbf{a}\right)$ The normalized score based on the average of Kendall rank correlation $\langle \tau \rangle$. $\left(\textbf{b}\right)$ The normalized score based on the standard deviation of Kendall rank correlation $S_{\tau}$. $\left(\textbf{c}\right)$ The normalized score based on the average of effective spreading coverage $\langle \tilde{s_n} \rangle$. Here, we show the $n=1,10,20$. The simulation results are obtained with a critical infection rate $\beta_{c}$.}
	\label{figure5}
\end{figure}

To validate the effectiveness of $RA$, we compare the $RA$ with other baseline methods on 40 disparate real networks according to the average of Kendall rank correlation $\langle \tau \rangle$, standard deviation of Kendall rank correlation $S_{\tau}$ and average of effective spreading coverage $\langle \tilde{s_n} \rangle$. Here, we employ the $NS$ to summarize the overall performance of different methods such that one gets a comprehensive understanding of the metrics' performance. The $\theta^*$ is determined in each network by the lowest standard deviation $S_{\tau}(\theta)$ instead of the largest $\langle \tau \rangle$, so the $RA(\theta^*)$ cannot guarantee the best performance of $RA$  among all parameters, which also means that its performance is more robust than other parameters under different $\lambda$. In Fig.\ref{figure5} $(a)$, we show the $NS$ of $\langle \tau \rangle$ for different methods. The overall performance of $RA(\theta^*)$ and $RA(\theta=2.5)$ are better than other methods in 40 real networks (The normalized value of $\langle \tau \rangle$ for different methods in each network are provided in \hyperref[sec: Appendix E]{Appendix E}). Since the $\theta^*$ is a tunable parameter and varies with the actual network, it is not competitive compared with other methods. Except for the $RA(*)$, $RA(\theta=2.5)$ has the best overall performance among all methods. Although the $NS$ of Subgraph/k is very close to the $RA(\theta=2.5)$, it considers global structural information. The $PI\_1$ and $PI\_2$ underperform than $RA(\theta=2.5)$ although they are proposed with the same idea. The possible reason is that the extent to penalize the high degree node could be controlled by parameter $\theta$ in $RA$. Therefore, the $RA$ with $\theta=2.5$ is a good metric to identify an effective spreader. Besides, the $NS$ of the standard deviation of $\tau$ for various methods in 40 networks is shown in Fig.\ref{figure5} $(b)$ (The normalized value of the standard deviation of $\tau$ for different methods in each network could be seen in  \hyperref[sec: Appendix E]{Appendix E}). For the $NS$ of $S_{\tau}$, the smaller the $NS$ is, the smaller the standard deviation of $\tau$. The Eigenvector/k has the smallest volatility among all methods under different $\lambda$. For the $RA(\theta=2.5)$, it could keep relative stability under different risk cases although it underperforms than the Eigenvector/k. Finally, the performance of $RA$ are discussed further according to the $\langle \tilde{s_n} \rangle$. Here, we identify the effective spreaders ranked in the top n=1, 10, and 20 by different methods and calculate the $\langle \tilde{s_n} \rangle$. The $NS$ of $\langle \tilde{s_n} \rangle$ for different methods is shown in Fig.\ref{figure5} $(c)$. The $NS$ of $RA(\theta^*)$ and $RA(2.5)$ for identifying the top 1, 10, 20 nodes are higher than other baselines methods. The overall performance of $RA(2.5)$ in 40 networks is better than other methods except for $RA(\theta^*)$, which again confirms that $RA(2.5)$ is a good predictor for identifying effective spreader.

\section{Conclusion}
%\section*{References}
As we all know,  a product promoted by a celebrity on a social network is rapidly spread to millions of users, while a similar product posted by a less connected individual cannot spread to many users. One of the most important factors determining the fate of the spreading process is where the initial spreader is located in the social network with a given connectivity pattern. So far, many methods have been proposed to identify the influential spreaders by utilizing purely the node topology features. However, when dealing with the real application of influence maximization, most of the existing studies have ignored the fact that it is more costly and difficult to convince influence nodes to act as initial spreaders, resulting in higher risk in maximizing the spreading. Therefore, we consider the real factors and introduce the activation risk of initial spreaders into the problem. In this paper, we assume that the probability of nodes to act as initial spreaders depend on their degree, and the large-degree nodes associate with lower activation probability than the small-degree nodes. For simplicity, we take the exponential decay function to map the degree of the node into the activated probability and introduce a risk parameter $\lambda$ to determine the difference of activation risk over various nodes. Through the theoretical result obtained from the percolation model on random networks, we find that the degree of the optimal initial spreader depends on the $\lambda$ and infection rate $\beta$ rather than the largest degree of nodes in the network. Meanwhile, we confirm the finding by numerical simulation on the ER network. In a real-world network,  we analyze that some existing centralities are correlated with degree. It might not enough to identify the effective spreaders by simply discounting for the degree in existing centralities. Thus, the risk-aware metric is proposed to identify effective spreaders. Experimental results about the normalized score of Kendall rank correlation $\langle \tau \rangle$ and the average of effective spreading coverage $\langle \tilde{s_n} \rangle$ on 40 disparate real networks show that our method outperforms some existing benchmark centralities. 

In actual problems, it is difficult to quantify the activation risk of influencers in marketing. The possibility to activate influencers depends on many factors, such as cost, brand awareness, and high-quality content. Among all factors, the cost of the influencer can be regarded as the main factor for successfully working with the influencer. Therefore, activation risk is very relevant to the cost of influencers. Due to the fact that the cost of influencers is an unsolved empirical problem, we introduce the activation risk to characterize the problem so that it fails to lose the generality, and assume that large-degree nodes tend to be activated with a lower probability. In a certain extent, the degree-decaying effect in effective spreading could be interpreted as higher cost. In addition, we employ the exponential decay function to describe the negative relation between the degree of node and activation risk because of analytic tractability and reasonable analytic results that agree with the intuition in reality. In fact, the conclusion obtained from results relies on the specific form of the activation function, which is proved in the \hyperref[sec:Appendix F]{Appendix F}. Although the analysis result is not robust against the form of activation function, the study still reveals that many existing centralities might not be enough to identify the effective spreaders once we consider the reality factors in node activation. For the practical problem, our findings suggest it is critical to propose new methods to identify effective spreaders that have a strong spreading ability but low degree because it could help us to design advertising and immunization strategy at a lower cost. The risk-aware metric can not only identify the effective spreaders but also be used to evaluate the potential importance of a node in the network. Meanwhile, it may also have additional "risk" to identify the most influential spreaders because it does not guarantee the outbreak size of detected nodes could achieve the maximization. We believe that this paper proposes a general research problem and many related issues could be studied in the near future. For example, different functional dependencies of effective spreading on the degree could be considered in the follow-up studies. One could design the effective spreading coverage on the basis of the middle-status conformity theory \cite{hu2014nonmonotonic} (i.e. individuals who are most likely to adopt an innovation or be susceptible to social contagion are those people in the middle strata of social status).

%A systematic study between the optimal strategy to initiate the spreading and network structure would be an interesting and important extension.

\section*{Acknowledgements}
This work is supported by the National Natural Science Foundation of China (Grant No.71731002), Fundamental Research Funds for the Central Universities with contract number 2019XD-A10 and National Key R\&D Program of China with contract number 2020YFF0305300.

\section*{Declaration of Competing Interest}
The authors declare that they have no known competing financial interests or personal relationships that could have appeared to influence the work reported in this paper.

\section*{Appendix A. The mean size of outbreak initiated from nodes with degree k}

\label{sec: Appendix A}

The mean size of the outbreak initiated from a randomly chosen node has been given by the bond percolation model and generation function\cite{newman2002spread}. Here, we wish to get the mean outbreak size of the disease $\langle s_k \rangle$ triggered from the single node with the degree $k$ on a random network. The ultimate size of the outbreak starting with a single infective node would be precisely the size of the cluster of nodes that can be reached from the initial node. In fact, the SIR model is equivalent to a bond percolation with bond occupation probability $T$. We employ the percolation model and generation function to give the exact mean size of the outbreak initiated from the single node with the degree $k$. 

Firstly, we need to define some generation functions because it could generate the probability distribution and easily work than probability distribution itself (The crucial properties of generation function could see the work \cite{PhysRevE.64.026118}). For instance, a generating function of degree distribution is as follow,
\begin{equation}
	G_0(x)=\sum_{k=0}^{\infty}{p_kx^k},
\end{equation}
the mean degree $\langle k \rangle$ of the node in the network is given by
\begin{equation}
	\langle k \rangle = \sum_{k}kp_{k}=G_0^{'}(1).
\end{equation}
If we follow an edge to the node at one of its ends, the probability of the reached nodes with degree $k$ is $q_k = \frac{kp_k}{\langle k \rangle}$. In general, we will concern with the number of ways of leaving such a node excluding the edge we arrived along, which is the degree minus 1. The distribution of degrees of the nodes reached by following a randomly chosen edge is generated by  
\begin{equation}
	G_1(x) = \sum_{k=1}^{\infty}\frac{kp_k}{\langle k \rangle}x^{k-1}=\frac{G_0^{'}(x)}{G_0^{'}(1)}.
\end{equation}
For a node with degree $k$,  the probability of node having exactly $m$ edges occupied from the $k$ edges follows the binomial distribution $\begin{pmatrix}k\\ m\end{pmatrix}T^m(1-T)^{k-m}$, 
hence the probability distribution of occupied edges for a node with degree $k$ is generated by Eqs.\ref{eqs11},
\begin{equation}
	G_{0}^{k}(x,T) =\sum_{m=0}^{\infty}\begin{pmatrix}k\\ m \end{pmatrix} T^m(1-T)^{k-m}x^m
	=\sum_{m=0}^{k}\begin{pmatrix} k\\ m \end{pmatrix}T^m(1-T)^{k-m}x^m
	=(1-T+xT)^k=(1-T(1-x))^k.
	\label{eqs11}
\end{equation}
Likewise, the probability distribution of occupied edges leaving the node arrived by following a randomly chosen edge is generated by Eqs.\ref{eqs12}
\begin{equation}
	G_{1}(x,T)=\sum_{m=0}^{\infty }\sum_{k=m}^{\infty}q_{k}\begin{pmatrix}
		k\\ m	
	\end{pmatrix}
	T^m(1-T)^{k-m}x^m = G_1(1-T(1-x)).
	\label{eqs12}
\end{equation}
We define the $\rho_s(T)$ as the distribution of cluster size of nodes reached by following a randomly chosen edge. Let $H_1(x,T)$ be the generating function for this distribution as is shown in the Eqs.\ref{eqs13},

\begin{equation}
	\label{eqs13}
	H_1(x,T)=\sum_{s=0}^{\infty}\rho_s(T) x^s.
\end{equation}
The $H_1$ can be broken down into an additive set of contributions as follows. We follow an edge to reach the cluster, which might be consisted by the following two parts: (1) a single node with no occupated edges connected to it other than the one along which we passed to reach it; (2) a single node with any number m ($m>1$) of occupated edge attached to it excluding the one along which we have reached it. Each occupated edge leading to another cluster whose size distribution is also generated by the $H_1$. The chance that a finite cluster containing a closed loop of edges goes as $N^{-1}$, and is zero in the limit of $N \to \infty$. Using these results, the $H_1(x,T)$ can be expressed in a Dyson-equation-like self-consistent form by Eqs.\ref{eqs14}, 
\begin{equation}
	\label{eqs14}
	H_1(x,T)=xG_1(H_1(x,T),T).
\end{equation}
The self-consistent process mentioned-above can be a better understanding by the following analysis. By analogy with the preceding part, then we represent the $p_s^k(T)$ as the distribution of the size of cluster reachable from a starting node with the degree $k$. We also give the generating function for this distribution in the Eqs.\ref{eqs15}. The $p_s^k(T)$ can be expressed by the $P(s-1|k)$, which means the probability of cluster size $s-1$ reached from the starting node with degree $k$, other than the starting node. The size of cluster $s$ can be broken down into different contributions from the $m$ occupied edges for a node with a degree $k$, which is described in the second row of Eqs.\ref{eqs15}. The $\delta$ is the Dirac delta function. If $s$ = $\sum_{r=1}^{m}j_r$, then $\delta = 1$, otherwise $\delta = 0$. The $\delta(s,\sum_{r=1}^{m}j_r)$ aims to hold that the sum of clustered size reached from the occupied edge is the same as s. The $\rho _{j_r}$ refers to the probability of cluster size $j_r$ of nodes reached by following a randomly chosen edge. By simplifying the equation, we obtain the form $h_0^k(x,T) = x(1-T(1-H_1(x,T)))^k$.
\begin{equation}
	\label{eqs15}
	\begin{aligned}
		& h_0^k(x,T)= \sum_{s=1}^{\infty}p_s^k(T) x^s= \sum_{s=1}^{\infty}P(s-1|k)(T) x^s = x\sum_{s=0}^{\infty}P(s|k)(T)x^s 
		\\ & = x\sum_{s=0}^{\infty}\sum_{m=0}^{k}\begin{pmatrix} k\\m  \end{pmatrix} T^m(1-T)^{k-m}\sum_{j_1}^{\infty }\cdots\sum_{j_m}^{\infty}\delta (s,\sum_{r=1}^{m}j_r)\prod _{r=1}^{m}\rho _{j_r}(T)x^s 
		\\& =x\sum_{m=0}^{k}\begin{pmatrix} k\\m \end{pmatrix} T^m(1-T)^{k-m}\sum_{s=0}^{\infty}\sum_{j_1}^{\infty }\cdots\sum_{j_m}^{\infty}\prod _{r=1}^{m}\rho _{j_r}(T)x^{j_r}
		\\ & =x\sum_{m=0}^{k}\begin{pmatrix} k\\m  \end{pmatrix} T^m(1-T)^{k-m}\left ( \sum_{j_r}^{\infty}{\rho_{j_r}(T)x^{j_r}} \right )^m 
		\\ & = x\sum_{m=0}^{k}\begin{pmatrix} k\\m  \end{pmatrix}(1-T)^{k-m}(T H_1(x,T))^m=x(1-T(1-H_1(x,T)))^k.
	\end{aligned}
\end{equation}

Then, we make full use of the important properties of generating function. The mean of the probability distribution is given by the first derivative of the generating function, evaluated at 1. Using the Eqs.\ref{eqs15}, we obtained the mean outbreak size of disease initiated from the node with degree $k$ by differentiating for $x$. At $x = 1$, we have
\begin{equation}
	\label{eqs16}
	\langle s_k \rangle = \frac{\partial h_0^k}{\partial x}\mid_{x=1}=[(1-T(1-H_1(x,T)))^k+xk(1-T(1-H_1(x,T)))^{k-1}T H_1^{'}(x,T)]|_{x=1}.
\end{equation}
By differentiating the Eqs.\ref{eqs14}, we have
\begin{equation}
	\label{eqs17}
	H_{1}^{'}(x,T) =G_1(H_1(x,T),T)+xG_{1}^{'}(H_1(x,T),T)H_{1}^{'}(x,T).
\end{equation}
By simplifying the Eqs.\ref{eqs17}, $H_{1}^{'}(x,T)$ would be 
\begin{equation}
	\label{eqs18}
	H_{1}^{'}(x,T) = \frac{G_1(H_1(x,T), T)}{1-xG_{1}^{'}[H_1(x,T),T]}.
\end{equation}
Due to the fact that the generating functions are 1 at $x =1$ if the distributions generated by generating functions are normalized, hence $H_1(1,T)=1$, $G_1(1,T)=1$. $G_1^{'}(x,T)=TG_1^{'}(1-T(1-x))$. Thus, we have 
\begin{equation}
	\label{eqs19}
	H_1^{'}(1,T) = \frac{1}{1-G_1^{'}(1,T)}=\frac{1}{1-TG_1^{'}(1)}.
\end{equation}
Substuting the Eqs.\ref{eqs19}, $H_1(1,T)=1$ and $G_1(1,T)=1$ into the Eqs.\ref{eqs16}, we obtain the exact value of mean outbreak size of disease triggered from the node with degree $k$, 
\begin{equation}
	\label{eqs20}
	\langle s_k\rangle = 1+\frac{kT}{1-TG_1^{'}(1)}.
\end{equation}
We take the derivative with respect to $G_1(x)$, it would be $G^{'}_1(x) = \sum_{k=2}^{\infty}{\frac{k(k-1)p_k}{\langle k \rangle}x^{k-2}}$, hence $G_1^{'}(1) = \frac{\langle k^2 \rangle- \langle k \rangle}{\langle k \rangle}$. We get the mean outbreak size of disease triggered from the node of degree k in closed form when $T_c < \frac{\langle k \rangle}{\langle k^2 \rangle- \langle k \rangle}$.

\section*{Appendix B. Network}
\label{sec: Network}
The risk-aware metric does not aim at specific datasets, so we examine the performance of our method on 40 datasets from different domains. Most of the datasets are downloaded from the network repository \footnote{\url{http://networkrepository.com/}}, other datasets are downloaded from the website \footnote{\url{http://networksciencebook.com/translations/en/resources/data.html}}. In the network repository, we randomly select 3 to 5 networks from different categories to form the corpus. We consider the largest connected component in each original network and analyze the statistical characteristic of networks. Detailed information see the Table \ref{table1}.

\captionsetup[table]{labelfont={bf},labelformat={default},labelsep=period,name={Table}}

\begin{table}[!htbp]
	\caption{\textbf{Structural properities on real networks.} Structural properities include the number of nodes in the network $\left(N\right)$, average degree $\left(\langle k \rangle\right)$, second order moment of degree($\langle k^2 \rangle$), maximum degree $\left(k_{max}\right)$, network diameter $\left( d \right)$, assortativity coefficient $\left( r \right)$ and critical infection rate $\left( \beta_{c} \right)$.} 
	\label{table1}
	\begin{tabular}{
			>{\columncolor[HTML]{FFFFFF}}c 
			>{\columncolor[HTML]{FFFFFF}}c 
			>{\columncolor[HTML]{FFFFFF}}c 
			>{\columncolor[HTML]{FFFFFF}}c 
			>{\columncolor[HTML]{FFFFFF}}c 
			>{\columncolor[HTML]{FFFFFF}}c 
			>{\columncolor[HTML]{FFFFFF}}c 
			>{\columncolor[HTML]{FFFFFF}}c }
		\toprule
		\cellcolor[HTML]{C0C0C0}{\color[HTML]{343434} }                                   & \cellcolor[HTML]{C0C0C0}{\color[HTML]{343434} }                             & \cellcolor[HTML]{C0C0C0}{\color[HTML]{343434} }                                                      & \cellcolor[HTML]{C0C0C0}{\color[HTML]{343434} }                                                                         & \cellcolor[HTML]{C0C0C0}{\color[HTML]{343434} }                                & \cellcolor[HTML]{C0C0C0}{\color[HTML]{343434} }                             & \cellcolor[HTML]{C0C0C0}{\color[HTML]{343434} }                             & \cellcolor[HTML]{C0C0C0}{\color[HTML]{343434} }                                \\
		\multirow{-2}{*}{\cellcolor[HTML]{C0C0C0}{\color[HTML]{343434} \textbf{$Networks$}}} & \multirow{-2}{*}{\cellcolor[HTML]{C0C0C0}{\color[HTML]{343434} \textbf{$N$}}} & \multirow{-2}{*}{\cellcolor[HTML]{C0C0C0}{\color[HTML]{343434} \textbf{\textless{}$k$\textgreater{}}}} & \multirow{-2}{*}{\cellcolor[HTML]{C0C0C0}{\color[HTML]{343434} \textbf{\textless{}$k^2$\textgreater{}}}} & \multirow{-2}{*}{\cellcolor[HTML]{C0C0C0}{\color[HTML]{343434} \textbf{$k_{max}$}}} & \multirow{-2}{*}{\cellcolor[HTML]{C0C0C0}{\color[HTML]{343434} \textbf{$d$}}} & \multirow{-2}{*}{\cellcolor[HTML]{C0C0C0}{\color[HTML]{343434} \textbf{$r$}}} & \multirow{-2}{*}{\cellcolor[HTML]{C0C0C0}{\color[HTML]{343434} \textbf{$\beta_c$}}} \\ \midrule
		\textit{ani-Mammalia}                                                             & 1430                                                                        & 5.45                                                                                                 & 48.88                                                                                                                   & 34                                                                             & 18                                                                          & 0.012                                                                       & 0.126                                                                          \\ 
		\textit{ani-Aves-Songbird}                                                        & 108                                                                         & 19                                                                                                   & 519.80                                                                                                                  & 56                                                                             & 6                                                                           & -0.005                                                                      & 0.038                                                                          \\ 
		\textit{ani-Reptilia}                                                             & 496                                                                         & 3.97                                                                                                 & 25.38                                                                                                                   & 17                                                                             & 21                                                                          & 0.345                                                                       & 0.185                                                                          \\ 
		\textit{ani-Dolphins}                                                             & 62                                                                          & 5.13                                                                                                 & 34.90                                                                                                                   & 12                                                                             & 8                                                                           & -0.044                                                                      & 0.172                                                                          \\ 
		\textit{bio-Celegans}                                                             & 453                                                                         & 8.94                                                                                                 & 358.49                                                                                                                  & 237                                                                            & 7                                                                           & -0.226                                                                      & 0.026                                                                          \\ 
		\textit{bio-Yeast}                                                                & 1458                                                                        & 2.67                                                                                                 & 19.05                                                                                                                   & 56                                                                             & 19                                                                          & -0.210                                                                      & 0.163                                                                          \\ 
		\textit{bio-Grid-Plant}                                                           & 1271                                                                        & 4.29                                                                                                 & 52.20                                                                                                                   & 71                                                                             & 26                                                                          & 0.001                                                                       & 0.090                                                                          \\ 
		\textit{bio-Grid-Worm}                                                            & 3342                                                                        & 3.85                                                                                                 & 196.54                                                                                                                  & 523                                                                            & 13                                                                          & -0.169                                                                      & 0.020                                                                          \\ 
		\textit{bn-Mouse-Kasthuri}                                                        & 986                                                                         & 3.12                                                                                                 & 50.87                                                                                                                   & 123                                                                            & 12                                                                          & -0.242                                                                      & 0.065                                                                          \\ 
		\textit{ca-CSphd}                                                                 & 1025                                                                        & 2.04                                                                                                 & 12.17                                                                                                                   & 46                                                                             & 28                                                                          & -0.253                                                                      & 0.201                                                                          \\ 
		\textit{ca-Erdos992}                                                              & 4991                                                                        & 2.98                                                                                                 & 48.83                                                                                                                   & 61                                                                             & 14                                                                          & -0.453                                                                      & 0.065                                                                          \\ 
		\textit{ca-GrQc}                                                                  & 4158                                                                        & 6.46                                                                                                 & 116.09                                                                                                                  & 81                                                                             & 17                                                                          & 0.639                                                                       & 0.059                                                                          \\ 
		\textit{ca-Netscience}                                                            & 379                                                                         & 4.82                                                                                                 & 38.69                                                                                                                   & 34                                                                             & 17                                                                          & -0.082                                                                      & 0.142                                                                          \\ 
		\textit{econ-Poli}                                                                & 2343                                                                        & 2.28                                                                                                 & 22.04                                                                                                                   & 63                                                                             & 27                                                                          & -0.335                                                                      & 0.115                                                                          \\ 
		\textit{econ-Mahindas}                                                            & 1258                                                                        & 12.03                                                                                                & 456.55                                                                                                                  & 206                                                                            & 8                                                                           & -0.060                                                                      & 0.027                                                                          \\ 
		\textit{econ-Wm1}                                                                 & 258                                                                         & 19.78                                                                                                & 945.28                                                                                                                  & 108                                                                            & 11                                                                          & -0.037                                                                      & 0.021                                                                          \\ 
		\textit{email-Dnc}                                                                & 1833                                                                        & 4.79                                                                                                 & 354.22                                                                                                                  & 404                                                                            & 8                                                                           & -0.305                                                                      & 0.014                                                                          \\ 
		\textit{email-Corecipient}                                                        & 849                                                                         & 24.46                                                                                                & 2276.19                                                                                                                 & 368                                                                            & 8                                                                           & -0.133                                                                      & 0.011                                                                          \\ 
		\textit{email-Enron-Only}                                                         & 143                                                                         & 8.71                                                                                                 & 112.59                                                                                                                  & 42                                                                             & 8                                                                           & -0.020                                                                      & 0.084                                                                          \\ 
		\textit{email-Univ}                                                               & 1133                                                                        & 9.62                                                                                                 & 179.82                                                                                                                  & 71                                                                             & 8                                                                           & 0.078                                                                       & 0.057                                                                          \\ 
		\textit{hs-Arenas-Jazz}                                                           & 198                                                                         & 27.70                                                                                                & 1070.24                                                                                                                 & 100                                                                            & 6                                                                           & 0.020                                                                       & 0.027                                                                          \\ 
		\textit{hs-Physical}                                                              & 117                                                                         & 7.95                                                                                                 & 79.16                                                                                                                   & 26                                                                             & 5                                                                           & -0.084                                                                      & 0.112                                                                          \\ 
		\textit{hs-Zachary}                                                               & 34                                                                          & 4.59                                                                                                 & 35.65                                                                                                                   & 17                                                                             & 5                                                                           & -0.476                                                                      & 0.148                                                                          \\ 
		\textit{ia-Crime-Moreno}                                                          & 829                                                                         & 3.55                                                                                                 & 21.69                                                                                                                   & 25                                                                             & 10                                                                          & -0.165                                                                      & 0.196                                                                          \\ 
		\textit{ia-Fb-Messages}                                                           & 1266                                                                        & 10.19                                                                                                & 279.09                                                                                                                  & 112                                                                            & 9                                                                           & -0.084                                                                      & 0.038                                                                          \\ 
		\textit{ia-Infect-Dublin}                                                         & 410                                                                         & 13.49                                                                                                & 252.43                                                                                                                  & 50                                                                             & 9                                                                           & 0.226                                                                       & 0.056                                                                          \\ 
		\textit{inf-Openflights}                                                          & 2905                                                                        & 10.77                                                                                                & 601.45                                                                                                                  & 242                                                                            & 14                                                                          & 0.049                                                                       & 0.018                                                                          \\ 
		\textit{inf-Euroroad}                                                             & 1039                                                                        & 2.51                                                                                                 & 7.75                                                                                                                    & 10                                                                             & 62                                                                          & 0.090                                                                       & 0.479                                                                          \\ 
		\textit{inf-Power}                                                                & 4941                                                                        & 2.67                                                                                                 & 10.33                                                                                                                   & 19                                                                             & 46                                                                          & 0.003                                                                       & 0.348                                                                          \\ 
		\textit{inf-Usair97}                                                              & 332                                                                         & 12.81                                                                                                & 568.16                                                                                                                  & 139                                                                            & 6                                                                           & -0.208                                                                      & 0.023                                                                          \\ 
		\textit{rt-Retweet}                                                               & 96                                                                          & 2.44                                                                                                 & 12.52                                                                                                                   & 17                                                                             & 10                                                                          & -0.179                                                                      & 0.241                                                                          \\ 
		\textit{rt-Twitter-Copen}                                                         & 761                                                                         & 2.70                                                                                                 & 22.22                                                                                                                   & 37                                                                             & 14                                                                          & -0.099                                                                      & 0.139                                                                          \\ 
		\textit{socfb-Caltech36}                                                          & 762                                                                         & 43.70                                                                                                & 3275.75                                                                                                                 & 248                                                                            & 6                                                                           & -0.066                                                                      & 0.014                                                                          \\ 
		\textit{socfb-Haverford76}                                                        & 1446                                                                        & 82.42                                                                                                & 10480.61                                                                                                                & 375                                                                            & 6                                                                           & 0.068                                                                       & 0.008                                                                          \\ 
		\textit{socfb-Reed98}                                                             & 962                                                                         & 39.11                                                                                                & 2784.14                                                                                                                 & 313                                                                            & 6                                                                           & 0.023                                                                       & 0.014                                                                          \\ 
		\textit{socfb-Simmons81}                                                          & 1510                                                                        & 43.69                                                                                                & 3197.12                                                                                                                 & 300                                                                            & 7                                                                           & -0.062                                                                      & 0.014                                                                          \\ 
		\textit{soc-Karate}                                                                   & 34                                                                          & 4.59                                                                                                 & 35.65                                                                                                                   & 17                                                                             & 5                                                                           & -0.476                                                                      & 0.148                                                                          \\ 
		\textit{Metabolic}                                                                & 1038                                                                        & 9.13                                                                                                 & 947.24                                                                                                                  & 637                                                                            & 6                                                                           & -0.250                                                                      & 0.010                                                                          \\ 
		\textit{Protein}                                                                  & 1646                                                                        & 3.06                                                                                                 & 35.71                                                                                                                   & 89                                                                             & 14                                                                          & -0.106                                                                      & 0.094                                                                          \\ 
		\textit{web-EPA}                                                                  & 4253                                                                        & 4.18                                                                                                 & 118.45                                                                                                                  & 175                                                                            & 10                                                                          & -0.304                                                                      & 0.037                                                                          \\ \bottomrule
	\end{tabular}
\end{table}

\section*{Appendix C. Correlation analysis}

\label{sec: Appendix C}

The effective spreading $\langle \tilde s_k \rangle$ is defined as $\langle s_k \rangle * p_k$ where the activation risk $p_k$ is quantified by the degree-decay function. In this way, some existing centrality metrics might underperform because they are correlated with the degree. To confirm the guess, we analyze Kendall rank correlation between degree and other centrality metrics. In fact, the correlation between two metrics depends on the network structure and varies with real networks. Here, we use two distinct ways to analyze the correlation in order to obtain a relative objective and comprehensive result on 40 networks. On one hand, we average the correlation between any two centrality metrics over 40 networks, which might have some bias due to the existence of extreme value. On the other hand, we count the ratio of the network in which Kendall rank correlation between two centralities is higher than a threshold (threshold = 0.5). If the results obtained from different method show the same phenomenon, a general conclusion could be drawn.

\begin{figure}[!htbp]
	\centering
	\includegraphics[width =1.0 \textwidth]{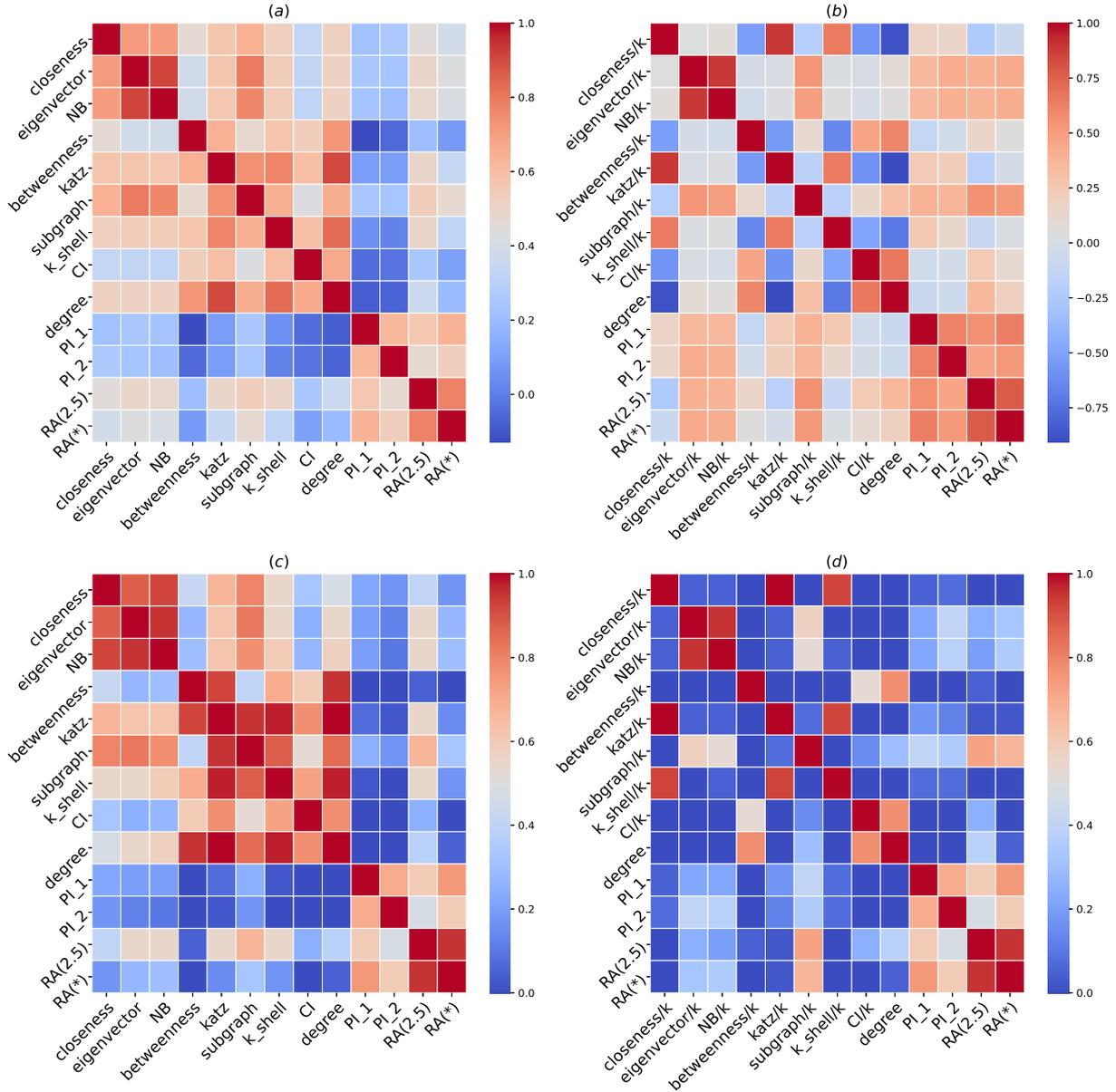}
	\caption{\textbf{The Kendall rank correlation between two centrality metrics on 40 networks.} The $RA(*)$ denotes that parameter $\theta$ is set as $\theta^*$ in each network where RA($\theta^*$) has the smallest volatility over different $\lambda$. $\left(\textbf{a}\right)$ The average of Kendall rank correlation between any two centralities. $\left(\textbf{b}\right)$ The average of Kendall rank correlation between two benchmark metrics. $\left(\textbf{c}\right)$ The proportion of network where Kendall rank correlation between two centralities is larger than $0.5$. $\left(\textbf{d}\right)$ The proportion of network where Kendall rank correlation between two benchmark centralities is larger than $0.5$.}
	\label{Sfigure1}
\end{figure}

The Kendall rank correlation between any two centrality metrics is analyzed. The results in Fig.\ref{Sfigure1} $(a)(c)$ show that centrality metrics could be roughly classified into three groups. The first group contains betweenness, Katz, subgraph, k-shell, and $CI$. The Kendall rank correlation between degree and them are very strong. The second group includes closeness, eigenvector, and $NB$. There is a weak correlation between them and degree. The third group includes $PI\_1$, $PI\_2$, $RA(2.5)$, and $RA(*)$, they are no correlated with a degree. Besides, the results show that centrality metrics within each group are correlated with each other, further confirming the reasonability of classification. Through the correlation analysis, we verify that it is not competitive to employ existing metrics as baseline methods to compare with the $RA$ method, because the activation probability in effective spreading has already considered the degree-decay effect and there is a strong correlation between degree and centralities in existing metrics, which naturally weakens the performance of existing metrics.

To eliminate the degree-decay effect, we calculate the ratio between centralities and degree (e.g. Katz score/degree, betweenness score/degree) as benchmark centralities. In Fig.\ref{Sfigure1} $(b)(d)$, the results show that there is no correlation between most benchmark centralities and degree. As for the benchmark centrality of closeness, Katz, and k-shell, they have a negative correlation with the degree, which suggests that making a comparison between them and $RA$ is more competitive when $\lambda$ is very large. As a result, it is relatively fair to compare $RA$ with other benchmark centralities by the effective spreading as a target function.

\section*{Appendix D. Statistical tests}

\label{sec: Appendix D}

Some rigorous statistical tests are provided to validate the result in the main text. We firstly make the significance tests of Kendall rank correlation between different methods and effective spreading coverage $\tilde{s}$. For various $\lambda$, the p-value for Kendall rank correlation coefficient is shown in Table \ref{table2}. The result corresponds to Fig.\ref{figure4} $(b)$. When $\lambda = 0.5$ and $0.6$, the Kendall rank correlation is not significant for some baseline methods, so the null hypothesis is supported, which suggests that there is no correlation between methods (BTN/k, CLO/k, Katz/k, KS/k, Degree) and effective spreading. In most cases, the baseline methods are correlated to effective spreading. Besides, we also test the Kendall rank correlation between $RA$ and effective spreading under different pair of parameters ($\lambda$ and $\beta$) in Table \ref{table3}, which corresponds to Fig.\ref{figure4} $(d)$. For most pairs of parameters, the p-value is lower than 0.05, which means the correlation between $RA$ and effective spreading coverage could be accepted. But for larger $\lambda$ and smaller $\theta$, or larger $\theta$ and smaller $\lambda$, there is no correlation between $RA$ and effective spreading coverage.  

For the Fig.\ref{figure4} $(c)$, some points among different methods might overlap due to the visual observation, we thus employ the Kolmogorov-Smirnov test (K-S test) to measure the difference of Kendall rank correlation distribution between $RA$ and other baseline methods under various $\lambda$, further verify that performance of our method is distinct with others. The K-S test is one of the most useful nonparametric methods to quantify a distance between the empirical distribution function of two samples. For a given infection rate $\beta$, two samples in the K-S test are respectively from Kendall rank correlation of both baseline method and $RA$ under different conditions. The p-value of the K-S test under different infection rates $\beta$ is shown in Table \ref{table4}. One could see that the p-value of the K-S test is equal to 0 between $RA$ and any baseline methods, suggesting that the distribution of Kendall correlation of $RA$ is statistically significantly different with other baseline methods.

\begin{table}[!htbp]
	\begin{tabular}{ccccccccccccc}
		\toprule
		\multicolumn{1}{l}{\cellcolor[HTML]{C0C0C0}}                                  & \cellcolor[HTML]{C0C0C0}                                        & \cellcolor[HTML]{C0C0C0}                                        & \cellcolor[HTML]{C0C0C0}                                        & \cellcolor[HTML]{C0C0C0}                                       & \cellcolor[HTML]{C0C0C0}                                         & \cellcolor[HTML]{C0C0C0}                                       & \cellcolor[HTML]{C0C0C0}                                       & \cellcolor[HTML]{C0C0C0}                                          & \cellcolor[HTML]{C0C0C0}                                      & \cellcolor[HTML]{C0C0C0}                                            & \cellcolor[HTML]{C0C0C0}                                          & \cellcolor[HTML]{C0C0C0}                                          \\
		\multicolumn{1}{l}{\multirow{-2}{*}{\cellcolor[HTML]{C0C0C0}\textbf{$\lambda$}}} & \multirow{-2}{*}{\cellcolor[HTML]{C0C0C0}\textit{\textbf{BTN/k}}} & \multirow{-2}{*}{\cellcolor[HTML]{C0C0C0}\textit{\textbf{CLO/k}}} & \multirow{-2}{*}{\cellcolor[HTML]{C0C0C0}\textit{\textbf{EIG/k}}} & \multirow{-2}{*}{\cellcolor[HTML]{C0C0C0}\textit{\textbf{NB/k}}} & \multirow{-2}{*}{\cellcolor[HTML]{C0C0C0}\textit{\textbf{Katz/k}}} & \multirow{-2}{*}{\cellcolor[HTML]{C0C0C0}\textit{\textbf{SG/k}}} & \multirow{-2}{*}{\cellcolor[HTML]{C0C0C0}\textit{\textbf{KS/k}}} & \multirow{-2}{*}{\cellcolor[HTML]{C0C0C0}\textit{\textbf{CI(3)/k}}} & \multirow{-2}{*}{\cellcolor[HTML]{C0C0C0}\textit{\textbf{Degree}}} & \multirow{-2}{*}{\cellcolor[HTML]{C0C0C0}\textit{\textbf{RA(2.5)}}} & \multirow{-2}{*}{\cellcolor[HTML]{C0C0C0}\textit{\textbf{PI\_1}}} & \multirow{-2}{*}{\cellcolor[HTML]{C0C0C0}\textit{\textbf{PI\_2}}} \\  \midrule 
		\textbf{0}                                                                    & 0                                                               & 0                                                               & 0                                                               & 0                                                              & 0                                                                &  0                                       & 0                                                              & 0                                                                 & 0                                                             & 0                                                                   & 0                                                                 & 0                                                                 \\ 
		\textbf{0.1}                                                                  & 0                                                               & 0                                                               & 0                                                               & 0                                                              & 0                                                                & 0                                                              & 0                                                              & 0                                                                 & 0                                                             & 0                                                                   & 0                                                                 & 0                                                                 \\ 
		\textbf{0.2}                                                                  & 0                                                               & 0                                                               & 0                                                               & 0                                                              & 0                                                                & 0                                                              & 0                                                              & 0                                                                 & 0                                                             & 0                                                                   & 0                                                                 & 0                                                                 \\ 
		\textbf{0.3}                                                                  & 0                                                               & 0                                                               & 0                                                               & 0                                                              & 0                                                                & 0                                                              & 0                                                              & 0                                                                 & 0                                                             & 0                                                                   & 0                                                                 & 0                                                                 \\ 
		\textbf{0.4}                                                                  & 0.0087                                                          & 0                                                               & 0                                                               & 0                                                              & 0                                                                & 0                                                              & 0.0322                                                         & 0                                                                 & 0                                                             & 0                                                                   & 0                                                                 & 0                                                                 \\ 
		\textbf{0.5}                                                                  & {\color[HTML]{9B9B9B} 0.6717}                                   & {\color[HTML]{9B9B9B} 0.1909}                                   & 0                                                               & 0                                                              & {\color[HTML]{9B9B9B} 0.2584}                                    & 0                                                              & {\color[HTML]{9B9B9B} 0.2862}                                  & 0                                                                 & 0.0001                                                        & 0                                                                   & 0                                                                 & 0                                                                 \\ 
		\textbf{0.6}                                                                  & 0.0013                                                          & 0.0266                                                          & 0                                                               & 0                                                              & 0.0131                                                           & 0                                                              & 0.0001                                                         & 0                                                                 & {\color[HTML]{9B9B9B} 0.7332}                                 & 0                                                                   & 0                                                                 & 0                                                                 \\ 
		\textbf{0.7}                                                                  & 0                                                               & 0                                                               & 0                                                               & 0                                                              & 0                                                                & 0                                                              & 0                                                              & 0                                                                 & 0.0041                                                        & 0                                                                   & 0                                                                 & 0                                                                 \\ 
		\textbf{0.8}                                                                  & 0                                                               & 0                                                               & 0                                                               & 0                                                              & 0                                                                & 0                                                              & 0                                                              & 0                                                                 & 0                                                             & 0                                                                   & 0                                                                 & 0                                                                 \\ 
		\textbf{0.9}                                                                  & 0                                                               & 0                                                               & 0                                                               & 0                                                              & 0                                                                & 0                                                              & 0                                                              & 0.0003                                                            & 0                                                             & 0                                                                   & 0                                                                 & 0                                                                 \\ 
		\textbf{1}                                                                    & 0                                                               & 0                                                               & 0                                                               & 0                                                              & 0                                                                & 0                                                              & 0                                                              & 0.0404                                                            & 0                                                             & 0                                                                   & 0                                                                 & 0                                                                 \\ \bottomrule
		
	\end{tabular}
	
	\caption{\textbf{The p-value for Kendall rank correlation between different methods and effective spreading $\tilde{s}$}. The number in grey denotes $p>0.05$, suggesting that the correlation is not significant. The full name of methods in the table is respectively betweenness ($BTN$), closeness ($CLO$), eigenvector ($EIG$), non-backtracking ($NB$), subgraph ($SG$), k-shell ($KS$), collective influence ($C I$), risk-aware metric ($RA(\theta=2.5)$), and potential influence ($PI\_1$ and $PI\_2$). The methods/k denotes the ratio between methods and degree.} 
	\label{table2}
\end{table}

% \usepackage{multirow}
\begin{table}[!htbp]
	\begin{tabular}{ccccccccccc}
		\toprule
		\rowcolor[HTML]{C0C0C0} 
		\cellcolor[HTML]{C0C0C0} & \cellcolor[HTML]{C0C0C0} & \cellcolor[HTML]{C0C0C0} & \cellcolor[HTML]{C0C0C0} & \cellcolor[HTML]{C0C0C0} & \cellcolor[HTML]{C0C0C0} & \cellcolor[HTML]{C0C0C0} & \cellcolor[HTML]{C0C0C0} & \cellcolor[HTML]{C0C0C0} & \cellcolor[HTML]{C0C0C0} & \cellcolor[HTML]{C0C0C0} \\
		\rowcolor[HTML]{C0C0C0} 
		\multirow{-2}{*}{\cellcolor[HTML]{C0C0C0}\textbf{$\lambda$}} & \multirow{-2}{*}{\cellcolor[HTML]{C0C0C0}\textit{\textbf{$\theta$=0}}} & \multirow{-2}{*}{\cellcolor[HTML]{C0C0C0}\textit{\textbf{$\theta$=0.5}}} & \multirow{-2}{*}{\cellcolor[HTML]{C0C0C0}\textit{\textbf{$\theta$=1}}} & \multirow{-2}{*}{\cellcolor[HTML]{C0C0C0}\textit{\textbf{$\theta$=1.5}}} & \multirow{-2}{*}{\cellcolor[HTML]{C0C0C0}\textit{\textbf{$\theta$=2}}} & \multirow{-2}{*}{\cellcolor[HTML]{C0C0C0}\textit{\textbf{$\theta$=2.5}}} & \multirow{-2}{*}{\cellcolor[HTML]{C0C0C0}\textit{\textbf{$\theta$=3}}} & \multirow{-2}{*}{\cellcolor[HTML]{C0C0C0}\textit{\textbf{$\theta$=3.5}}} & \multirow{-2}{*}{\cellcolor[HTML]{C0C0C0}\textit{\textbf{$\theta$=4}}} & \multirow{-2}{*}{\cellcolor[HTML]{C0C0C0}\textit{\textbf{$\theta$=4.5}}} \\ \midrule 
		
		\textbf{0} & 0 & 0 & 0 & 0 & 0 & 0 & 0 & 0 & 0 & 0 \\ 
		\textbf{0.1} & 0 & 0 & 0 & 0 & 0 & 0 & 0 & 0 & 0 & 0 \\ 
		\textbf{0.2} & 0 & 0 & 0 & 0 & 0 & 0 & 0 & 0 & 0 & 0 \\ 
		\textbf{0.3} & 0 & 0 & 0 & 0 & 0 & 0 & 0 & 0 & 0 & 0 \\ 
		\textbf{0.4} & 0 & 0 & 0 & 0 & 0 & 0 & 0 & 0 & 0 & 0 \\ 
		\textbf{0.5} & 0.0001 & 0 & 0 & 0 & 0 & 0 & 0 & 0 & 0 & 0 \\ 
		\textbf{0.6} & {\color[HTML]{C0C0C0} 0.7332} & 0.0027 & 0 & 0 & 0 & 0 & 0 & 0 & 0 & 0 \\ 
		\textbf{0.7} & 0.0041 & {\color[HTML]{C0C0C0} 0.8605} & 0.0163 & 0 & 0 & 0 & 0 & 0 & 0 & 0 \\ 
		\textbf{0.8} & 0 & 0.0038 & {\color[HTML]{C0C0C0} 0.7904} & 0 & 0 & 0 & 0 & 0 & 0 & 0 \\ 
		\textbf{0.9} & 0 & 0 & 0.0113 & 0.0345 & 0 & 0 & 0 & 0 & 0 & 0 \\ 
		\textbf{1.0} & 0 & 0 & 0 & {\color[HTML]{C0C0C0} 0.9913} & 0 & 0 & 0 & 0 & 0 & 0 \\  \bottomrule
		\toprule
		\cellcolor[HTML]{C0C0C0} & \cellcolor[HTML]{C0C0C0} & \cellcolor[HTML]{C0C0C0} & \cellcolor[HTML]{C0C0C0} & \cellcolor[HTML]{C0C0C0} & \cellcolor[HTML]{C0C0C0} & \cellcolor[HTML]{C0C0C0} & \cellcolor[HTML]{C0C0C0} & \cellcolor[HTML]{C0C0C0} & \cellcolor[HTML]{C0C0C0} & \cellcolor[HTML]{C0C0C0} \\
		\multirow{-2}{*}{\cellcolor[HTML]{C0C0C0}\textit{\textbf{$\lambda$}}} & \multirow{-2}{*}{\cellcolor[HTML]{C0C0C0}\textit{\textbf{$\theta$=5}}} & \multirow{-2}{*}{\cellcolor[HTML]{C0C0C0}\textit{\textbf{$\theta$=5.5}}} & \multirow{-2}{*}{\cellcolor[HTML]{C0C0C0}\textit{\textbf{$\theta$=6}}} & \multirow{-2}{*}{\cellcolor[HTML]{C0C0C0}\textit{\textbf{$\theta$=6.5}}} & \multirow{-2}{*}{\cellcolor[HTML]{C0C0C0}\textit{\textbf{$\theta$=7}}} & \multirow{-2}{*}{\cellcolor[HTML]{C0C0C0}\textit{\textbf{$\theta$=7.5}}} & \multirow{-2}{*}{\cellcolor[HTML]{C0C0C0}\textit{\textbf{$\theta$=8}}} & \multirow{-2}{*}{\cellcolor[HTML]{C0C0C0}\textit{\textbf{$\theta$=8.5}}} & \multirow{-2}{*}{\cellcolor[HTML]{C0C0C0}\textit{\textbf{$\theta$=9}}} & \multirow{-2}{*}{\cellcolor[HTML]{C0C0C0}\textit{\textbf{$\theta$=9.5}}} \\ 
		\midrule
		\textbf{0} & 0.001 & 0.0109 & {\color[HTML]{C0C0C0} 0.0626} & {\color[HTML]{C0C0C0} 0.2026} & {\color[HTML]{C0C0C0} 0.4421} & {\color[HTML]{C0C0C0} 0.6468} & {\color[HTML]{C0C0C0} 0.9066} & {\color[HTML]{C0C0C0} 0.9002} & {\color[HTML]{C0C0C0} 0.7454} & {\color[HTML]{C0C0C0} 0.637} \\ 
		\textbf{0.1} & 0 & 0.0001 & 0.0018 & 0.0113 & 0.0425 & {\color[HTML]{C0C0C0} 0.0864} & {\color[HTML]{C0C0C0} 0.1689} & {\color[HTML]{C0C0C0} 0.2572} & {\color[HTML]{C0C0C0} 0.3513} & {\color[HTML]{C0C0C0} 0.4373} \\ 
		\textbf{0.2} & 0 & 0 & 0 & 0 & 0.0002 & 0.0006 & 0.0019 & 0.0042 & 0.0077 & 0.0121 \\ 
		\textbf{0.3} & 0 & 0 & 0 & 0 & 0 & 0 & 0 & 0 & 0 & 0 \\ 
		\textbf{0.4} & 0 & 0 & 0 & 0 & 0 & 0 & 0 & 0 & 0 & 0 \\ 
		\textbf{0.5} & 0 & 0 & 0 & 0 & 0 & 0 & 0 & 0 & 0 & 0 \\ 
		\textbf{0.6} & 0 & 0 & 0 & 0 & 0 & 0 & 0 & 0 & 0 & 0 \\ 
		\textbf{0.7} & 0 & 0 & 0 & 0 & 0 & 0 & 0 & 0 & 0 & 0 \\ 
		\textbf{0.8} & 0 & 0 & 0 & 0 & 0 & 0 & 0 & 0 & 0 & 0 \\ 
		\textbf{0.9} & 0 & 0 & 0 & 0 & 0 & 0 & 0 & 0 & 0 & 0 \\ 
		\textbf{1.0} & 0 & 0 & 0 & 0 & 0 & 0 & 0 & 0 & 0 & 0 \\ \bottomrule
	\end{tabular}
	\caption{\textbf{The p-value of Kendall rank correlation between $RA$ and effective spreading $\tilde{s}$ under various $\lambda$ and $\theta$.} The number in grey denotes $p > 0.05$, suggesting that the correlation is not significant.} 
	\label{table3}
\end{table}

% \usepackage{multirow}
% \usepackage{multirow}
\begin{table}[!htbp]
	\begin{tabular}{cccccccccc}
		\toprule
		\cellcolor[HTML]{C0C0C0} & \cellcolor[HTML]{C0C0C0} & \cellcolor[HTML]{C0C0C0} & \cellcolor[HTML]{C0C0C0} & \cellcolor[HTML]{C0C0C0} & \cellcolor[HTML]{C0C0C0} & \cellcolor[HTML]{C0C0C0} & \cellcolor[HTML]{C0C0C0} & \cellcolor[HTML]{C0C0C0} & \cellcolor[HTML]{C0C0C0} \\
		\multirow{-2}{*}{\cellcolor[HTML]{C0C0C0}\textbf{$\beta$}} & \multirow{-2}{*}{\cellcolor[HTML]{C0C0C0}\textit{\textbf{K}}} & \multirow{-2}{*}{\cellcolor[HTML]{C0C0C0}\textit{\textbf{KS/k}}} & \multirow{-2}{*}{\cellcolor[HTML]{C0C0C0}\textit{\textbf{BTN/k}}} & \multirow{-2}{*}{\cellcolor[HTML]{C0C0C0}\textit{\textbf{CLO/k}}} & \multirow{-2}{*}{\cellcolor[HTML]{C0C0C0}\textit{\textbf{EIG/k}}} & \multirow{-2}{*}{\cellcolor[HTML]{C0C0C0}\textit{\textbf{KZ/k}}} & \multirow{-2}{*}{\cellcolor[HTML]{C0C0C0}\textit{\textbf{SG/k}}} & \multirow{-2}{*}{\cellcolor[HTML]{C0C0C0}\textit{\textbf{NB/k}}} & \multirow{-2}{*}{\cellcolor[HTML]{C0C0C0}\textit{\textbf{CI(3)/k}}} \\ \midrule
		\textbf{0.02} & 0 & 0 & 0 & 0 & 0 & 0 & 0 & 0 & 0 \\ 
		\textbf{0.04} & 0 & 0 & 0 & 0 & 0 & 0 & 0 & 0 & 0 \\ 
		\textbf{0.06} & 0 & 0 & 0 & 0 & 0 & 0 & 0 & 0 & 0 \\ 
		\textbf{0.08} & 0 & 0 & 0 & 0 & 0 & 0 & 0 & 0 & 0 \\ 
		\textbf{0.10} & 0 & 0 & 0 & 0 & 0 & 0 & 0 & 0 & 0 \\ 
		\textbf{0.12} & 0 & 0 & 0 & 0 & 0 & 0 & 0 & 0 & 0 \\ 
		\textbf{0.14} & 0 & 0 & 0 & 0 & 0 & 0 & 0 & 0 & 0 \\ 
		\textbf{0.16} & 0 & 0 & 0 & 0 & 0 & 0 & 0 & 0 & 0 \\ 
		\textbf{0.18} & 0 & 0 & 0 & 0 & 0 & 0 & 0 & 0 & 0 \\ 
		\textbf{0.20} & 0 & 0 & 0 & 0 & 0 & 0 & 0 & 0 & 0 \\ \bottomrule
	\end{tabular}
	\caption{\textbf{The p-value of Kolmogorov-Smirnov test for the difference of two empirical distribution under distinct infection rate $\beta$, i.e. $RA(\theta=2.5)$ and other baseline methods.} The full name of methods in the table is respectively degree($K$), k-shell($KS$), betweenness ($BTN$), closeness ($CLO$), eigenvector ($EIG$), Katz ($KZ$), subgraph ($SG$), non-backtracking ($NB$), and collective influence($CI$). The methods/k denotes the ratio between methods and degree.} 
	\label{table4}
\end{table}

\section*{Appendix E. Normalized score}
\label{sec: Appendix E}

According to different evaluation metrics, we normalize the performance of different methods on each network, further obtaining the overall performance on all networks. The normalized value of Kendall rank correlation between benchmark centralities and effective spreading could be seen in Table \ref{Table5}. Besides, we also show the normalized value of the standard deviation of Kendall rank correaltion between centralities and effective spreading coverage in Table \ref{Table6}.

% \usepackage{booktabs}
% \usepackage{multirow}
\begin{table}[]
	\setlength{\tabcolsep}{1mm}{
		\begin{tabular}{@{}cccccccccccccc@{}}
			\toprule
			\cellcolor[HTML]{C0C0C0} & \cellcolor[HTML]{C0C0C0} & \cellcolor[HTML]{C0C0C0} & \cellcolor[HTML]{C0C0C0} & \cellcolor[HTML]{C0C0C0} & \cellcolor[HTML]{C0C0C0} & \cellcolor[HTML]{C0C0C0} & \cellcolor[HTML]{C0C0C0} & \cellcolor[HTML]{C0C0C0} & \cellcolor[HTML]{C0C0C0} & \cellcolor[HTML]{C0C0C0} & \cellcolor[HTML]{C0C0C0} & \cellcolor[HTML]{C0C0C0} & \cellcolor[HTML]{C0C0C0} \\
			\multirow{-2}{*}{\cellcolor[HTML]{C0C0C0}\textit{\textbf{Networks}}} & \multirow{-2}{*}{\cellcolor[HTML]{C0C0C0}\textit{\textbf{BTN/k}}} & \multirow{-2}{*}{\cellcolor[HTML]{C0C0C0}\textit{\textbf{CLO/k}}} & \multirow{-2}{*}{\cellcolor[HTML]{C0C0C0}\textit{\textbf{EIG/k}}} & \multirow{-2}{*}{\cellcolor[HTML]{C0C0C0}\textit{\textbf{NB/k}}} & \multirow{-2}{*}{\cellcolor[HTML]{C0C0C0}\textit{\textbf{Katz/k}}} & \multirow{-2}{*}{\cellcolor[HTML]{C0C0C0}\textit{\textbf{SG/k}}} & \multirow{-2}{*}{\cellcolor[HTML]{C0C0C0}\textit{\textbf{KS/k}}} & \multirow{-2}{*}{\cellcolor[HTML]{C0C0C0}\textit{\textbf{CI(3)/k}}} & \multirow{-2}{*}{\cellcolor[HTML]{C0C0C0}\textit{\textbf{Degree}}} & \multirow{-2}{*}{\cellcolor[HTML]{C0C0C0}\textit{\textbf{PI\_1}}} & \multirow{-2}{*}{\cellcolor[HTML]{C0C0C0}\textit{\textbf{PI\_2}}} & \multirow{-2}{*}{\cellcolor[HTML]{C0C0C0}\textit{\textbf{RA(2.5)}}} & \multirow{-2}{*}{\cellcolor[HTML]{C0C0C0}\textit{\textbf{RA(*)}}} \\ \midrule
			\textit{email-Univ} & 0.69 & 0.00 & 0.85 & 0.84 & 0.00 & 0.99 & 0.11 & 0.85 & 0.88 & 0.74 & 0.80 & \textbf{1.00} & 0.94 \\
			\textit{ani-Aves-Songbird} & 0.57 & 0.00 & 0.87 & 0.86 & 0.00 & 0.98 & 0.11 & 0.12 & 0.82 & 0.59 & 0.59 & \textbf{1.00} & \textbf{1.00} \\
			\textit{ani-Dolphins} & 0.60 & 0.03 & 0.91 & 0.89 & 0.00 & 0.86 & 0.02 & 0.61 & 0.77 & 0.55 & 0.43 & \textbf{1.00} & \textbf{1.00} \\
			\textit{ani-Reptilia} & 0.42 & 0.03 & 0.79 & 0.79 & 0.00 & 0.90 & 0.13 & 0.87 & 0.70 & 0.68 & 0.90 & \textbf{1.00} & 0.94 \\
			\textit{ani-Mammalia} & 0.47 & 0.00 & 0.76 & 0.72 & 0.01 & \textbf{1.00} & 0.15 & 0.94 & 0.68 & 0.73 & 0.83 & 0.95 & 0.93 \\
			\textit{bio-Celegans} & 0.18 & 0.57 & \textbf{1.00} & \textbf{1.00} & 0.50 & 0.78 & 0.63 & 0.00 & 0.22 & 0.95 & 0.84 & 0.64 & 0.95 \\
			\textit{bio-Grid-Plant} & 0.00 & 0.31 & 0.81 & 0.62 & 0.35 & 0.99 & 0.40 & 0.32 & 0.15 & \textbf{1.00} & 0.95 & 0.87 & 0.98 \\
			\textit{bio-Grid-Worm} & 0.07 & 0.99 & 0.69 & 0.67 & \textbf{1.00} & 0.61 & 0.89 & 0.07 & 0.00 & 0.69 & 0.68 & 0.43 & 0.69 \\
			\textit{bio-Yeast} & 0.07 & 0.51 & 0.86 & 0.87 & 0.51 & 0.86 & 0.34 & 0.23 & 0.00 & \textbf{1.00} & 0.97 & 0.93 & 0.93 \\
			\textit{bn-Mouse-Kasthuri} & 0.17 & 0.72 & 0.83 & 0.89 & 0.73 & 0.86 & 0.71 & 0.28 & 0.00 & 0.95 & 0.98 & 0.77 & \textbf{1.00} \\
			\textit{ca-CSphd} & 0.06 & 0.62 & 0.70 & 0.44 & 0.88 & 0.93 & 0.61 & 0.07 & 0.00 & \textbf{1.00} & 0.94 & 0.95 & 0.91 \\
			\textit{ca-Erdos992} & 0.00 & 0.68 & 0.88 & 0.87 & 0.67 & \textbf{1.00} & 0.48 & 0.15 & 0.07 & 0.94 & 0.89 & 0.78 & 0.96 \\
			\textit{ca-GrQc} & 0.06 & 0.02 & 0.97 & 0.93 & 0.00 & \textbf{1.00} & 0.11 & 0.71 & 0.26 & 0.91 & 0.95 & 0.83 & 0.94 \\
			\textit{ca-Netscience} & 0.08 & 0.00 & 0.66 & 0.61 & 0.01 & 0.99 & 0.11 & 0.66 & 0.31 & 0.94 & 0.94 & \textbf{1.00} & \textbf{1.00} \\
			\textit{econ-Mahindas} & 0.56 & 0.00 & 0.97 & 0.97 & 0.01 & \textbf{1.00} & 0.32 & 0.65 & 0.89 & 0.98 & 0.89 & 0.99 & 0.90 \\
			\textit{econ-Poli} & 0.07 & 0.71 & 0.55 & 0.58 & 0.95 & 0.96 & 0.69 & 0.08 & 0.00 & \textbf{1.00} & 0.94 & 0.89 & 0.89 \\
			\textit{econ-Wm1} & 0.35 & 0.02 & 0.99 & 0.99 & 0.00 & 0.94 & 0.25 & 0.24 & 0.78 & 0.90 & 0.39 & \textbf{1.00} & 0.92 \\
			\textit{email-Dnc} & 0.09 & 0.99 & 0.70 & 0.68 & \textbf{1.00} & 0.50 & 0.87 & 0.10 & 0.00 & 0.55 & 0.63 & 0.30 & 0.49 \\
			\textit{email- Corecipient} & 0.27 & 0.02 & 0.57 & 0.57 & 0.18 & 0.85 & 0.07 & 0.00 & 0.30 & \textbf{1.00} & 0.69 & 0.60 & 0.94 \\
			\textit{email-Enron-Only} & 0.48 & 0.01 & 0.90 & 0.89 & 0.00 & 0.98 & 0.21 & 0.46 & 0.81 & 0.67 & 0.76 & \textbf{1.00} & \textbf{1.00} \\
			\textit{hs-Arenas-Jazz} & 0.53 & 0.01 & 0.90 & 0.89 & 0.00 & \textbf{1.00} & 0.23 & 0.11 & 0.88 & 0.55 & 0.60 & 0.98 & 0.98 \\
			\textit{hs-Physical} & 0.56 & 0.00 & 0.81 & 0.77 & 0.01 & 0.96 & 0.07 & 0.10 & 0.72 & 0.47 & 0.50 & 0.87 & \textbf{1.00} \\
			\textit{hs-Zachary} & 0.50 & 0.28 & 0.65 & 0.60 & 0.31 & 0.72 & 0.34 & 0.00 & 0.45 & 0.93 & 0.65 & 0.99 & \textbf{1.00} \\
			\textit{ia-Crime-Moreno} & 0.64 & 0.02 & 0.97 & 0.94 & 0.00 & 0.51 & 0.09 & 0.84 & 0.58 & 0.81 & 0.86 & \textbf{1.00} & \textbf{1.00} \\
			\textit{ia-Fb-Messages} & 0.75 & 0.00 & 0.80 & 0.79 & 0.00 & 0.96 & 0.13 & 0.71 & 0.87 & 0.99 & 0.91 & \textbf{1.00} & 0.99 \\
			\textit{ia-Infect-Dublin} & 0.51 & 0.04 & 0.86 & 0.86 & 0.00 & \textbf{1.00} & 0.23 & 0.80 & 0.87 & 0.53 & 0.59 & \textbf{1.00} & 0.95 \\
			\textit{inf-Euroroad} & 0.06 & 0.16 & 0.99 & \textbf{1.00} & 0.00 & 0.31 & 0.08 & 0.95 & 0.37 & 0.34 & 0.59 & 0.73 & 0.81 \\
			\textit{inf-Openflights} & 0.16 & 0.00 & 0.87 & 0.87 & 0.00 & 0.87 & 0.29 & 0.40 & 0.34 & 0.99 & 0.86 & 0.71 & \textbf{1.00} \\
			\textit{inf-Power} & 0.21 & 0.01 & \textbf{1.00} & 0.25 & 0.09 & 0.52 & 0.00 & 0.75 & 0.32 & 0.59 & 0.76 & 0.87 & 0.90 \\
			\textit{inf-Usair97} & 0.30 & 0.00 & 0.70 & 0.70 & 0.00 & 0.87 & 0.22 & 0.05 & 0.50 & \textbf{1.00} & 0.78 & 0.79 & 0.98 \\
			\textit{Metabolic} & 0.26 & \textbf{1.00} & 0.72 & 0.71 & 0.98 & 0.46 & 0.88 & 0.54 & 0.00 & 0.44 & 0.66 & 0.24 & 0.44 \\
			\textit{Protein} & 0.00 & 0.38 & 0.85 & 0.88 & 0.39 & 0.96 & 0.33 & 0.22 & 0.01 & \textbf{1.00} & 0.93 & 0.86 & 0.97 \\
			\textit{rt-Retweet} & 0.00 & 0.49 & \textbf{1.00} & 0.98 & 0.47 & 0.75 & 0.13 & 0.17 & 0.08 & 0.98 & \textbf{1.00} & \textbf{1.00} & \textbf{1.00} \\
			\textit{rt-Twitter} & 0.00 & 0.49 & \textbf{1.00} & \textbf{1.00} & 0.46 & 0.95 & 0.24 & 0.17 & 0.00 & \textbf{1.00} & 0.94 & 0.87 & 0.95 \\
			\textit{soc-Karate} & 0.20 & 0.14 & 0.46 & 0.44 & 0.16 & 0.62 & 0.28 & 0.00 & 0.33 & 0.87 & 0.50 & 0.96 & \textbf{1.00} \\
			\textit{socfb-Caltech36} & 0.74 & 0.00 & 0.75 & 0.75 & 0.04 & \textbf{1.00} & 0.11 & 0.16 & 0.97 & 0.70 & 0.62 & \textbf{1.00} & \textbf{1.00} \\
			\textit{socfb-Haverford76} & 0.72 & 0.02 & 0.88 & 0.88 & 0.01 & \textbf{1.00} & 0.11 & 0.00 & 0.93 & 0.58 & 0.49 & 0.98 & 0.98 \\
			\textit{socfb-Reed98} & 0.60 & 0.00 & 0.83 & 0.82 & 0.07 & 0.97 & 0.17 & 0.06 & 0.90 & 0.73 & 0.72 & \textbf{1.00} & 0.97 \\
			\textit{socfb-Simmons81} & 0.80 & 0.00 & 0.74 & 0.74 & 0.03 & \textbf{1.00} & 0.06 & 0.16 & 0.96 & 0.58 & 0.55 & 0.94 & 0.87 \\
			\textit{web-EPA} & 0.11 & 0.96 & 0.81 & 0.77 & \textbf{1.00} & 0.73 & 0.88 & 0.11 & 0.00 & 0.85 & 0.86 & 0.43 & 0.83 \\ \bottomrule
	\end{tabular}}
	\caption{\textbf{Normalized value of Kendall rank correlation between methods and effective spreading in all networks.} The name of methods in the table is the same as the Table.\ref{table2}. The $RA(*)$ denotes that $\theta$ is determined by the smallest standard deviation of Kendall rank correlation over different $\lambda$ in each network. The number in bold denotes the method performs well than other methods in the current network.}
\label{Table5}
\end{table}

% \usepackage{booktabs}
% \usepackage{multirow}
\begin{table}[]
	\setlength{\tabcolsep}{1mm}{
		\begin{tabular}{@{}clllllllllllll@{}}
			\toprule
			\cellcolor[HTML]{C0C0C0} & \multicolumn{1}{c}{\cellcolor[HTML]{C0C0C0}} & \multicolumn{1}{c}{\cellcolor[HTML]{C0C0C0}} & \multicolumn{1}{c}{\cellcolor[HTML]{C0C0C0}} & \multicolumn{1}{c}{\cellcolor[HTML]{C0C0C0}} & \multicolumn{1}{c}{\cellcolor[HTML]{C0C0C0}} & \multicolumn{1}{c}{\cellcolor[HTML]{C0C0C0}} & \multicolumn{1}{c}{\cellcolor[HTML]{C0C0C0}} & \multicolumn{1}{c}{\cellcolor[HTML]{C0C0C0}} & \multicolumn{1}{c}{\cellcolor[HTML]{C0C0C0}} & \multicolumn{1}{c}{\cellcolor[HTML]{C0C0C0}} & \multicolumn{1}{c}{\cellcolor[HTML]{C0C0C0}} & \multicolumn{1}{c}{\cellcolor[HTML]{C0C0C0}} & \multicolumn{1}{c}{\cellcolor[HTML]{C0C0C0}} \\
			\multirow{-2}{*}{\cellcolor[HTML]{C0C0C0}\textit{\textbf{Networks}}} & \multicolumn{1}{c}{\multirow{-2}{*}{\cellcolor[HTML]{C0C0C0}\textit{\textbf{BTN/k}}}} & \multicolumn{1}{c}{\multirow{-2}{*}{\cellcolor[HTML]{C0C0C0}\textit{\textbf{CLO/k}}}} & \multicolumn{1}{c}{\multirow{-2}{*}{\cellcolor[HTML]{C0C0C0}\textit{\textbf{EIG/k}}}} & \multicolumn{1}{c}{\multirow{-2}{*}{\cellcolor[HTML]{C0C0C0}\textit{\textbf{NB/k}}}} & \multicolumn{1}{c}{\multirow{-2}{*}{\cellcolor[HTML]{C0C0C0}\textit{\textbf{Katz/k}}}} & \multicolumn{1}{c}{\multirow{-2}{*}{\cellcolor[HTML]{C0C0C0}\textit{\textbf{SG/k}}}} & \multicolumn{1}{c}{\multirow{-2}{*}{\cellcolor[HTML]{C0C0C0}\textit{\textbf{KS/k}}}} & \multicolumn{1}{c}{\multirow{-2}{*}{\cellcolor[HTML]{C0C0C0}\textit{\textbf{CI(3)/k}}}} & \multicolumn{1}{c}{\multirow{-2}{*}{\cellcolor[HTML]{C0C0C0}\textit{\textbf{Degree}}}} & \multicolumn{1}{c}{\multirow{-2}{*}{\cellcolor[HTML]{C0C0C0}\textit{\textbf{PI\_1}}}} & \multicolumn{1}{c}{\multirow{-2}{*}{\cellcolor[HTML]{C0C0C0}\textit{\textbf{PI\_2}}}} & \multicolumn{1}{c}{\multirow{-2}{*}{\cellcolor[HTML]{C0C0C0}\textit{\textbf{RA(2.5)}}}} & \multicolumn{1}{c}{\multirow{-2}{*}{\cellcolor[HTML]{C0C0C0}\textit{\textbf{RA(*)}}}} \\ \midrule
			\textit{email-Univ} & 0.68 & 0.96 & 0.03 & \textbf{0.00} & 0.97 & 0.56 & 0.98 & 0.34 & 1.00 & 0.52 & 0.13 & 0.45 & 0.07 \\
			\textit{ani-Aves-Songbird} & 0.55 & 0.98 & \textbf{0.00} & 0.02 & 0.98 & 0.57 & 0.99 & 0.53 & 1.00 & 0.80 & 0.60 & 0.07 & 0.07 \\
			\textit{ani-Dolphins} & 0.46 & 0.92 & 0.04 & 0.13 & 0.95 & 0.60 & 0.96 & \textbf{0.00} & 1.00 & 0.78 & 0.81 & 0.17 & 0.17 \\
			\textit{ani-Reptilia} & 0.42 & 0.91 & 0.02 & \textbf{0.00} & 0.92 & 0.33 & 0.64 & 0.54 & 1.00 & 0.59 & 0.29 & 0.29 & 0.06 \\
			\textit{ani-Mammalia} & 0.68 & 0.93 & \textbf{0.00} & \textbf{0.00} & 0.95 & 0.53 & 0.95 & 0.71 & 1.00 & 0.63 & 0.29 & 0.17 & 0.12 \\
			\textit{bio-Celegans} & 0.69 & 0.92 & 0.12 & 0.11 & 0.94 & 0.55 & 0.76 & 0.30 & 1.00 & 0.36 & \textbf{0.00} & 0.65 & 0.19 \\
			\textit{bio-Grid-Plant} & 0.59 & 0.85 & 0.14 & \textbf{0.00} & 0.87 & 0.26 & 0.81 & 0.66 & 1.00 & 0.30 & 0.10 & 0.35 & 0.14 \\
			\textit{bio-Grid-Worm} & 0.73 & 0.77 & \textbf{0.00} & 0.02 & 0.77 & 0.19 & 0.59 & 0.88 & 1.00 & 0.15 & 0.09 & 0.48 & 0.11 \\
			\textit{bio-Yeast} & 0.74 & 0.79 & 0.01 & \textbf{0.00} & 0.79 & 0.05 & 0.88 & 0.88 & 1.00 & 0.31 & 0.08 & 0.06 & 0.06 \\
			\textit{bn-Mouse-Kasthuri} & 0.80 & 0.79 & 0.06 & \textbf{0.00} & 0.79 & 0.09 & 0.76 & 0.80 & 1.00 & 0.14 & 0.09 & 0.22 & 0.10 \\
			\textit{ca-CSphd} & 0.88 & 0.69 & 0.01 & \textbf{0.00} & 0.69 & 0.46 & 0.95 & 0.89 & 1.00 & 0.33 & 0.12 & 0.19 & 0.13 \\
			\textit{ca-Erdos992} & 0.86 & 0.63 & 0.01 & \textbf{0.00} & 0.63 & 0.18 & 0.78 & 0.95 & 1.00 & 0.19 & 0.05 & 0.37 & 0.13 \\
			\textit{ca-GrQc} & 0.42 & 0.92 & 0.06 & \textbf{0.00} & 0.93 & 0.23 & 0.50 & 0.57 & 1.00 & 0.08 & 0.03 & 0.51 & 0.01 \\
			\textit{ca-Netscience} & 0.77 & 0.90 & \textbf{0.00} & 0.01 & 0.93 & 0.54 & 0.84 & 0.46 & 1.00 & 0.67 & 0.32 & 0.20 & 0.20 \\
			\textit{econ-Mahindas} & 0.16 & 0.95 & 0.41 & 0.41 & 0.95 & 0.55 & 0.44 & 0.26 & 1.00 & 0.14 & 0.08 & 0.68 & \textbf{0.00} \\
			\textit{econ-Poli} & 0.85 & 0.68 & \textbf{0.00} & 0.01 & 0.69 & 0.14 & 0.89 & 0.90 & 1.00 & 0.23 & 0.13 & 0.16 & 0.16 \\
			\textit{econ-Wm1} & \textbf{0.00} & 0.92 & 0.46 & 0.45 & 0.93 & 0.72 & 0.80 & 0.16 & 1.00 & 0.19 & 0.45 & 0.62 & 0.07 \\
			\textit{email-Dnc} & 0.73 & 0.74 & \textbf{0.00} & 0.01 & 0.74 & 0.38 & 0.62 & 0.72 & 1.00 & 0.36 & 0.14 & 0.67 & 0.42 \\
			\textit{email- Corecipient} & 0.48 & 0.93 & 0.05 & 0.04 & 0.73 & 0.45 & 0.63 & 0.10 & 1.00 & 0.07 & 0.02 & 0.78 & \textbf{0.00} \\
			\textit{email-Enron-Only} & 0.53 & 0.94 & 0.11 & 0.09 & 0.97 & 0.62 & 0.87 & \textbf{0.00} & 1.00 & 0.70 & 0.28 & 0.17 & 0.17 \\
			\textit{hs-Arenas-Jazz} & 0.55 & 0.98 & \textbf{0.00} & \textbf{0.00} & 0.99 & 0.62 & 0.85 & 0.42 & 1.00 & 0.77 & 0.59 & 0.09 & 0.09 \\
			\textit{hs-Physical} & 0.70 & 0.95 & \textbf{0.00} & 0.19 & 0.95 & 0.44 & 0.99 & 0.34 & 1.00 & 0.86 & 0.35 & 0.42 & 0.03 \\
			\textit{hs-Zachary} & 0.73 & 0.89 & 0.41 & 0.54 & 0.91 & 0.31 & 0.95 & \textbf{0.00} & 1.00 & 0.51 & 0.73 & 0.27 & 0.12 \\
			\textit{ia-Crime-Moreno} & 0.77 & 0.90 & \textbf{0.00} & 0.23 & 0.90 & \textbf{0.00} & 0.99 & 0.87 & 1.00 & 0.60 & 0.24 & 0.11 & 0.11 \\
			\textit{ia-Fb-Messages} & 0.74 & 0.96 & \textbf{0.00} & 0.04 & 0.96 & 0.75 & 0.99 & 0.08 & 1.00 & 0.19 & 0.11 & 0.66 & 0.12 \\
			\textit{ia-Infect-Dublin} & 0.31 & 0.96 & \textbf{0.00} & \textbf{0.00} & 0.98 & 0.28 & 0.80 & 0.06 & 1.00 & 0.69 & 0.38 & 0.24 & 0.07 \\
			\textit{inf-Euroroad} & 0.45 & 0.77 & 0.01 & \textbf{0.00} & 0.79 & 0.28 & 0.82 & 0.70 & 1.00 & 0.69 & 0.43 & 0.17 & 0.06 \\
			\textit{inf-Openflights} & 0.61 & 0.92 & 0.11 & 0.11 & 0.92 & 0.47 & 0.80 & 0.34 & 1.00 & 0.11 & 0.34 & 0.70 & \textbf{0.00} \\
			\textit{inf-Power} & 0.62 & 0.86 & 0.12 & \textbf{0.00} & 0.87 & 0.30 & 0.91 & 0.80 & 1.00 & 0.73 & 0.46 & 0.25 & 0.14 \\
			\textit{inf-Usair97} & 0.74 & 0.96 & \textbf{0.00} & 0.01 & 0.96 & 0.60 & 0.83 & 0.03 & 1.00 & 0.28 & 0.04 & 0.80 & 0.22 \\
			\textit{Metabolic} & 0.32 & 0.91 & 0.08 & 0.08 & 0.93 & 0.51 & 0.65 & 0.52 & 1.00 & 0.58 & \textbf{0.00} & 0.75 & 0.52 \\
			\textit{Protein} & 0.73 & 0.82 & 0.12 & 0.08 & 0.82 & 0.09 & 0.84 & 0.87 & 1.00 & 0.03 & 0.07 & 0.20 & \textbf{0.00} \\
			\textit{rt-Retweet} & 0.84 & 0.78 & 0.02 & 0.19 & 0.78 & \textbf{0.00} & 0.93 & 0.78 & 1.00 & 0.41 & 0.07 & 0.05 & 0.05 \\
			\textit{rt-Twitter} & 0.82 & 0.78 & 0.06 & 0.02 & 0.78 & \textbf{0.00} & 0.93 & 0.91 & 1.00 & 0.28 & 0.09 & 0.22 & 0.05 \\
			\textit{soc-Karate} & 0.75 & 0.89 & 0.29 & 0.49 & 0.89 & 0.26 & 1.00 & 0.02 & 0.99 & 0.47 & 0.70 & 0.22 & \textbf{0.00} \\
			\textit{socfb-Caltech36} & 0.64 & 0.99 & \textbf{0.00} & 0.03 & 0.93 & 0.87 & 0.97 & 0.82 & 1.00 & 0.61 & 0.38 & 0.12 & 0.12 \\
			\textit{socfb-Haverford76} & 0.62 & 1.00 & 0.02 & \textbf{0.00} & 1.00 & 0.76 & 0.97 & 0.79 & 1.00 & 0.67 & 0.61 & 0.01 & 0.01 \\
			\textit{socfb-Reed98} & 0.70 & 0.99 & 0.01 & \textbf{0.00} & 0.94 & 0.84 & 0.96 & 0.80 & 1.00 & 0.58 & 0.07 & 0.34 & 0.10 \\
			\textit{socfb-Simmons81} & 0.66 & 0.99 & 0.02 & \textbf{0.00} & 0.93 & 0.79 & 0.98 & 0.65 & 1.00 & 0.60 & 0.23 & 0.16 & 0.09 \\
			\textit{web-EPA} & 0.80 & 0.83 & \textbf{0.00} & 0.01 & 0.83 & 0.21 & 0.82 & 0.91 & 1.00 & 0.09 & 0.01 & 0.60 & 0.08 \\ \bottomrule
	\end{tabular}}
	\caption{\textbf{Normalized value of the standard deviation of Kendall rank correlation between methods and effective spreading}. The name of methods in the table is the same as the Table.\ref{table2}. The $RA(*)$ denotes that $\theta$ is determined by the smallest standard deviation of Kendall rank correlation among all parameters $\theta$ ($\theta \in [0,9.5]$ with a step of 0.5). The number in bold denotes that Kendall rank correlation of the method has the smallest volatility under different $\lambda$.}
\label{Table6}
\end{table}

\section*{Appendix F. Activation function}
\label{sec:Appendix F}

\begin{figure}[!htbp]
	\centering
	\includegraphics[width =1.0 \textwidth]{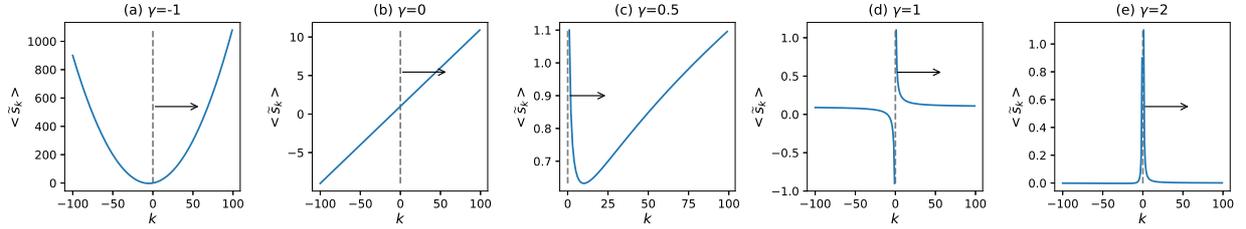}
	\caption{\textbf{The effective spreading $\langle \tilde{s_k} \rangle$ for various $\gamma$}, $ \langle \tilde{s_k} \rangle = k^{-\gamma}+k^{1-\gamma}q, q=\frac{\beta\beta_c}{\beta_c-\beta}$. Here $q$ is set as $0.1$. One could clearly see that the change of function form as the increase of $\gamma$. The $k$ value corresponding to largest effective spreading coverage on the right side of dashed line is the optimal degree of node in the problem.}  
	\label{Sfigure2}
\end{figure}

In defined problem, the choice of activation function is important to analytic tractability, and the optimal initial spreaders about the problem also depends on the slection of activation function. To provide evidence to support the claim,  we choose the $\frac{1}{k^{\gamma}}$ as activation function to make a further analysis. $p_k=\frac{1}{k^{\gamma}}, \gamma$ is a parameter to control the risk difference among nodes with different degree. Firstly, we substitute  $\frac{1}{k^{\gamma}}$  into the $\langle \tilde{s_k} \rangle = p_k * s_k$, see the Eqs.\ref{eqs27}. 
	\begin{equation}
		\langle \tilde{s_k} \rangle = \frac{1}{k^{\gamma}}(1+\frac{k\beta}{1-\frac{\beta}{\beta_c}})=k^{-\gamma} + k^{1-\gamma}\frac{\beta\beta_c}{\beta_c -\beta}.
		\label{eqs27}
	\end{equation}
Then, we derive analytic solution by setting $\frac{\partial \langle \tilde{s_k} \rangle}{\partial k}=0$, $k^* = (\frac{1}{\beta}-\frac{1}{\beta_c})(\frac{\gamma}{1-\gamma}),\beta < \beta_c$. (1) When $-1 \leq \gamma < 0$, $k^*$ is less than 0 and is the minimal point of $\langle \tilde{s_k} \rangle$, which suggests that the degree of the optimal initial spreaders in the problem should be the largest degree among all nodes in the network (see the Fig.\ref{Sfigure2} $(a)$). (2) When $\gamma = 0$, then $\langle \tilde{s_k} \rangle = 1 + \frac{k\beta}{1-\frac{\beta}{\beta_c}}$, the problem degrenerates into the original problem (see the Fig.\ref{Sfigure2} $(b)$) and the degree of the optimal initial spreaders is the largest degree among all nodes in the network. (3) When $0 < \gamma < 1$, $k^*$ is greater than $0$ and corresponds to the minimal point of $\langle \tilde{s_k} \rangle$ (see the Fig.\ref{Sfigure2} $(c)$). The degree of the optimal initial spreaders in the problem is difficult to be determined. By calculating partial derivatives for $\gamma$, $\frac{\partial k^*}{\partial \gamma}=(\frac{1}{\beta}-\frac{1}{\beta_c})(\frac{1}{1-\gamma})^2>0$, we find that $k^*$ increases as $\gamma$, which suggests that the minimal point of $\langle \tilde{s_k} \rangle$ tends to be large as the increase of $\gamma$, further inferring that the degree of the optimal initial spreaders should be the largest degree among all nodes when $\gamma$ is close to 0 and the smallest degree among all nodes when $\gamma$ is close to 1. (4) When $\gamma=1$, $\langle \tilde{s_k} \rangle = k^{-1}+\frac{\beta\beta_c}{\beta_c-\beta}$ (see the Fig.\ref{Sfigure2} $(d)$), which suggests the degree of the optimal initial spreaders should be the smallest degree among all nodes in the network. In summary, the degree of the optimal initial spreaders to maximize the effective spreading $\langle \tilde{s_k} \rangle$ corresponds to the largest degree among all nodes in network when $\gamma < \gamma_c$, and corresponds to the smallest degree among all nodes in network when $\gamma > \gamma_c$ ($\gamma_c$ is a threshold). The results here are qualitatively similar to the exponential decay function, namely the degree of the optimal initial spreaders have a negative relation with $\gamma$.

\begin{figure}[!htbp]
	\centering
	\includegraphics[width =1.0 \textwidth]{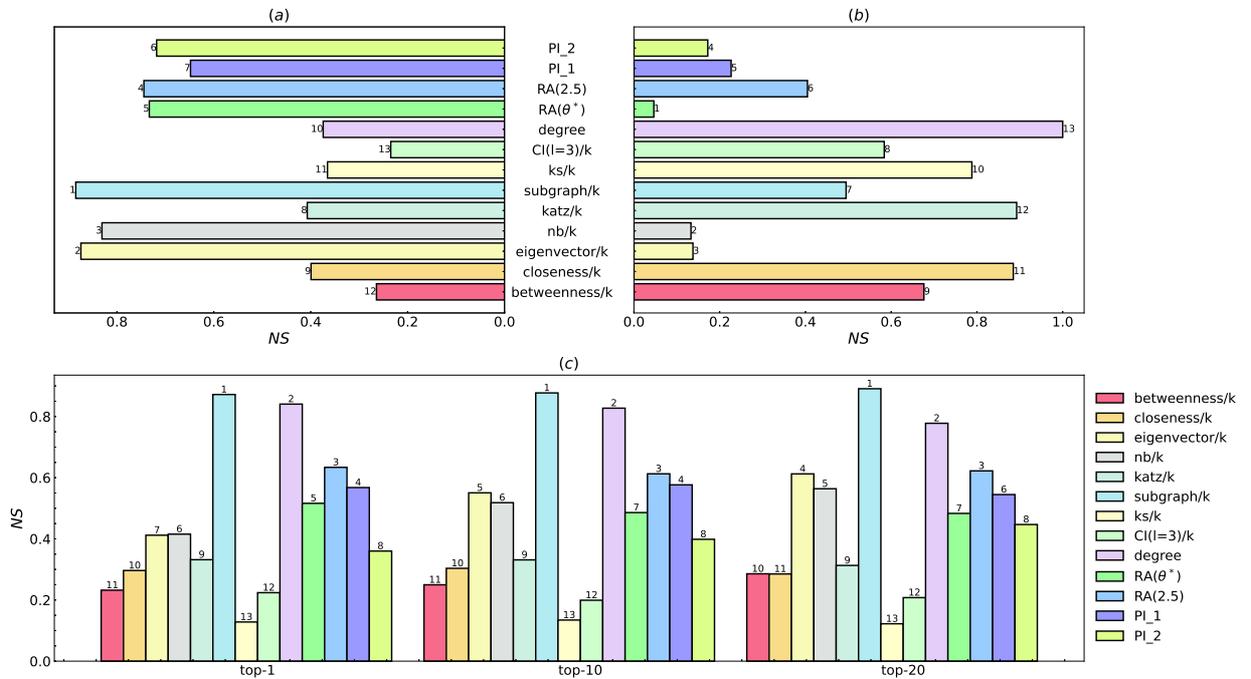}
	\caption{\textbf{Based on the activation function of $\frac{1}{k^\gamma}$ , the normalized score of evaluation metrics for different methods in all networks}. $\left(\textbf{a}\right)$ The normalized score of average of Kendall $\tau$. $\left(\textbf{b}\right)$ The normalized score of standard deviation of Kendall $\tau$. $\left(\textbf{c}\right)$ The normalized score of average of effective spreading coverage $\langle \tilde{s_n} \rangle$.} 
	\label{Sfigure3}
\end{figure}

\begin{figure}[!htbp]
	\centering
	\includegraphics[width =1.0 \textwidth]{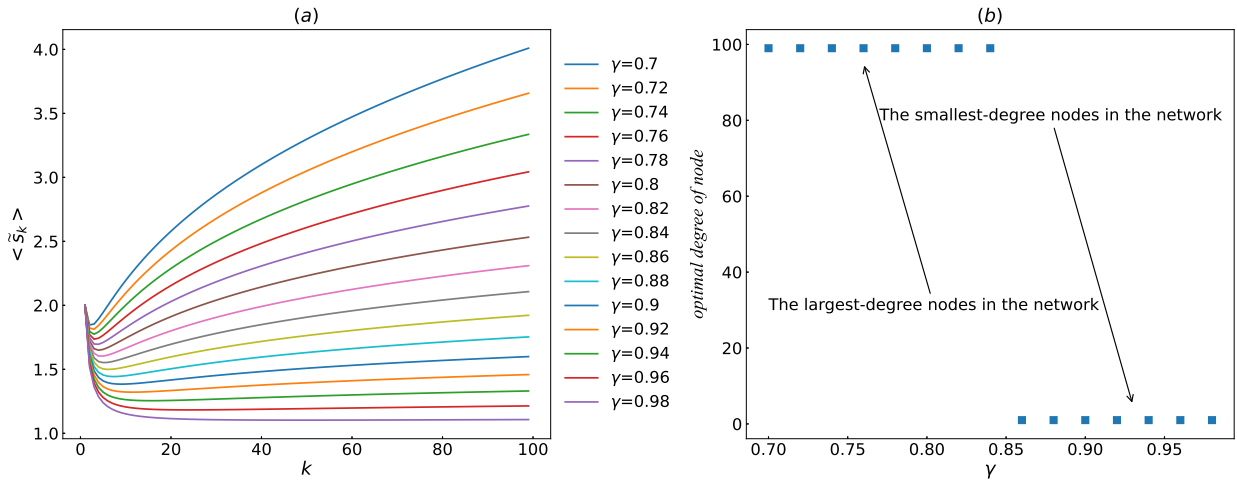}
	\caption{ $\left(\textbf{a}\right)$ The $\langle \tilde{s_k} \rangle$ plotted as a function of $k$ for different $\gamma$, $\gamma \in [0.7,0.98]$ and $q =1$. $\left(\textbf{b}\right)$ The optimal initial spreaders to maximize the effective spreading for different $\gamma$. Here, we assume that the network size is 100.} 
	\label{Sfigure4}	
\end{figure}

We conduct the experiment to test the performance of risk-aware metric according to the activation function ($\frac{1}{k^\gamma}$). $\gamma$ is set from 0 to 1 with a step of 0.1. The setting of $\gamma$ is the same as the risk parameter $\lambda$. The experimental result is shown in Fig.\ref{Sfigure3}. One could see that the performance of $RA(\theta=2.5)$ for $\langle \tau \rangle$, $S_{\tau}$ and $\langle \tilde{s} \rangle$ is poorer than other baseline methods. The $RA(\theta^*)$ is used to make a comparison, it lacks an advantage. The possible reason behind the results is that the optimal initial spreaders for most of $\gamma$ in interval $\gamma \in [0,1]$ favor the largest-degree nodes in the network, and the optimal initial spreaders for less $\gamma$ favor the smallest-degree nodes. In other words, the degree of the optimal spreaders is not continuous when we consider $\frac{1}{k^\gamma}$ as an activation function, which could be confirmed by the simulation result. In the Eqs.\ref{eqs27}, the effective spreading $\langle \tilde{s_k} \rangle = k^{-\gamma}+k^{1-\gamma} q,q=\frac{\beta\beta_c}{\beta_c-\beta}$. When $q$ is given, the $\langle \tilde{s_k} \rangle$ plotted as a function of $k$ for different $\gamma$ could be seen in the Fig.\ref{Sfigure4} $(a)$. One could clearly observe the change of effective spreading as the increase of $\gamma$. In Fig.\ref{Sfigure4} $(b)$, the degree of the optimal initial spreaders for various $\gamma$ shows the discontinuous transition. The condition where the degree of the optimal spreaders is in the intermediate degree does not exist, which is different from the analytical result of exponential decay activation. When we select $\frac{1}{k^\gamma}$ as an activation function, the largest effective spreading coverage among all nodes does not correspond to the nodes linked to many hubs but the largest-degree node or smallest-degree node in the real networks. This could explain why the $RA$ has poorer performance than other metrics, and degree and subgraph benchmark centrality have better performance.

Through the above analysis, the inversely proportional function might not be a good form as activation function because it is hard to analyze the degree of the optimal initial spreaders and there is no exact analytical solution for the optimal degree value in the defined problem. Intuitively, the optimal initial spreaders obtained from the $\frac{1}{k^\gamma}$ is not agreed well with the real condition, although it could characterize the negative relation between the degree and activation probability. Therefore, the form of the activation function is important to the quantification of the problem. The conclusion obtained from the result depends on the activation function.

\bibliography{mybibfile}
\end{document}